\documentclass[fleqn,usenatbib]{mnras}

\usepackage{newtxtext,newtxmath}

\usepackage[T1]{fontenc}
\usepackage{ae,aecompl}

\usepackage{graphicx}	
\usepackage{amsmath}	
\usepackage{amssymb}	
\usepackage[dvipsnames]{xcolor}

\newcommand{\oocrit}{\,\Omega/\Omega_{\rm crit,\,ZAMS}}

\title[Stellar rotation in NGC~1846]{How stellar rotation shapes the colour magnitude diagram of the massive intermediate-age star cluster NGC~1846}

\author[S. Kamann et al.]{%
S. Kamann,$^{1}$\thanks{E-mail: s.kamann@ljmu.ac.uk}
N. Bastian,$^{1}$
S. Gossage,$^{2}$
D. Baade,$^{3}$
I. Cabrera-Ziri,$^{2}$\thanks{Hubble Fellow}
\newauthor
G. Da Costa,$^{4}$
S. E. de Mink,$^{2,5}$
C. Georgy,$^{6}$  
B. Giesers,$^{7}$
F. G\"ottgens,$^{7}$
\newauthor
M. Hilker,$^{3}$  
T.-O. Husser,$^{7}$
C. Lardo,$^{8}$
S.~S. Larsen,$^{9}$  
D. Mackey,$^{4}$
S. Martocchia,$^{1,3}$
\newauthor
A. Mucciarelli,$^{10,11}$
I. Platais,$^{12}$
M.~M. Roth,$^{13}$
M. Salaris,$^{1}$
C. Usher,$^{1}$
D. Yong$^{4}$
\\
$^{1}$Astrophysics Research Institute, Liverpool John Moores University, 146 Brownlow Hill, Liverpool L3 5RF, UK\\
$^{2}$Harvard \& Smithsonian Center for Astrophysics, 60 Garden Street, Cambridge, MA 02138, USA\\
$^{3}$ESO, European Southern Observatory, Karl-Schwarzschild-Str. 2, 85748 Garching bei M\"unchen, Germany\\
$^{4}$Research School of Astronomy and Astrophysics, Australian National University, Canberra, ACT 0200, Australia\\
$^{5}$Anton Pannekoek Institute for Astronomy and GRAPPA, University of Amsterdam, Science Park 904, 1098~XH Amsterdam, The Netherlands\\
$^{6}$Department of Astronomy, University of Geneva, Chemin des Maillettes 51, 1290 Versoix, Switzerland\\
$^{7}$Institute for Astrophysics, Georg-August-University, Friedrich-Hund-Platz 1, 37077 G\"ottingen, Germany\\
$^{8}$Laboratoire d'astrophysique, \'Ecole Polytechnique F\'ed\'erale de Lausanne (EPFL), Observatoire de Sauverny, CH-1290 Versoix, CH\\
$^{9}$Department of Astrophysics/IMAPP, Radboud University, P.O. Box 9010, 6500~GL Nijmegen, The Netherlands\\
$^{10}$INAF-Osservatorio di Astrofisica \& Scienza dello Spazio, via Gobetti 93/3, I-40129, Bologna, Italy\\
$^{11}$Dipartimento di Fisica \& Astronomia, Universit\` a degli Studi di Bologna, via Gobetti 93/2, I-40129, Bologna, Italy\\
$^{12}$Department of Physics and Astronomy, Johns Hopkins 
University, 3400 North Charles Street, Baltimore, MD 21218, USA\\
$^{13}$Leibniz-Institute for Astrophysics, An der Sternwarte 16, 14482 Potsdam, Germany
}

\date{Accepted XXX. Received YYY; in original form ZZZ}

\pubyear{2019}

\begin{document}
\label{firstpage}
\pagerange{\pageref{firstpage}--\pageref{lastpage}}
\maketitle

\begin{abstract}
We present a detailed study of stellar rotation in the massive 1.5~Gyr old cluster NGC~1846 in the Large Magellanic Cloud. Similar to other clusters at this age, NGC~1846 shows an extended main sequence turn-off (eMSTO), and previous photometric studies have suggested it could be bimodal. In this study, we use MUSE integral-field spectroscopy to measure the projected rotational velocities ($v\sin i$) of around $1\,400$ stars across the eMSTO and along the upper main sequence of NGC~1846. We measure $v\sin i$ values up to $\sim250\,{\rm km\,s^{-1}}$ and find a clear relation between the $v\sin i$ of a star and its location across the eMSTO. Closer inspection of the distribution of rotation rates reveals evidence for a bimodal distribution, with the fast rotators centred around $v\sin i=140\,{\rm km\,s^{-1}}$ and the slow rotators centred around $v\sin i=60\,{\rm km\,s^{-1}}$. We further observe a lack of fast rotating stars along the photometric binary sequence of NGC~1846, confirming results from the field that suggest that tidal interactions in binary systems can spin down stars. However, we do not detect a significant difference in the binary fractions of the fast and slowly rotating sub-populations. Finally, we report on the serendipitous discovery of a planetary nebula associated with NGC~1846.
\end{abstract}




\section{Introduction}

It is now well established that the resolved colour-magnitude diagrams (CMDs) of young and intermediate-age stellar clusters (i.e. $<2~{\rm Gyr}$) display features that are not well reproduced through simple stellar isochrones, such as extended main sequence turn-offs \citep[eMSTOs, e.g.,][]{2007MNRAS.379..151M} and split or dual main-sequences \citep[e.g.][]{2015MNRAS.450.3750M}.  These features have been suggested to be due to distributions of stellar rotational velocities \citep[e.g.][]{2009MNRAS.398L..11B,2015ApJ...807...25B,2019arXiv190711251G}. Rotation changes the internal structure of the star, because the centrifugal support and extra mixing in the core region alter its hydrostatic equilibrium compared to that of a non-rotating star of the same mass and composition. These effects modify the evolutionary path of the rotating star in the CMD relative to the equivalent non-rotating star \citep[e.g.][]{2000A&A...361..101M}. Furthermore, in stars rotating close to their break-up velocity, the spherical symmetry is broken, leading to effective temperature variations across its surface \citep{1924MNRAS..84..665V}, so that the observed magnitudes also become a function of the inclination of the rotation axis towards the line of sight. Finally, the additional mixing can change the surface abundances of a star compared to the non-rotating case. However, the relative impact of these effects is still a matter of debate and varies across stellar evolutionary models. Further, we note that rotation is not the only possible process affecting the hydrostatic equilibrium of a star \citep[see][]{2019arXiv191000591J}.

Several studies argued that the impact of stellar rotation on the CMDs would be minor \citep[e.g.,][]{2011MNRAS.412L.103G,2012ApJ...751L...8P} and that other factors such as age spreads or differential extinction would be responsible for the observed features. However, recent observations have directly linked the position of stars in the CMDs to their projected rotational velocity ($v\sin i$) values \citep[e.g.][]{2017ApJ...846L...1D,2018MNRAS.480.1689K,2018MNRAS.480.3739B,2018AJ....156..116M}.

In order to explain features such as dual main sequences, it has been suggested that the rotational distribution of stars in young clusters may also be bimodal, with peaks at near zero and near critical (i.e. break-up) velocity \citep[e.g.][]{2015MNRAS.453.2637D}. Evidence for the presence of stars rotating close to their break-up velocities comes from the discovery of large populations of Be stars in young clusters \citep{2017MNRAS.465.4795B}. While rotational velocities of field stars with similar masses and spectral types show an extended and perhaps bimodal distribution \citep[see][for early A and late B stars]{2007A&A...463..671R}, this is not extreme enough to match what is needed to explain the observed oddities in the CMDs of young massive clusters. In addition, \citet{2007A&A...463..671R} found that the bimodality in the distribution of rotational velocities among Galactic field stars disappears for intermediate A-stars and later spectral types, where, aside from members of binary systems, only fast rotating stars are observed. This raises the question of what causes the large range of rotational velocities required to explain the eMSTOs of intermediate-age clusters. \citet{2015MNRAS.453.2637D} suggested that binary stars could play a role. Braking via tidal interactions appears to work efficiently in field binary stars, which \citet{2004ApJ...616..562A} found to be predominantly slow rotators. However, in some clusters, the slowly rotating stars appear to be more numerous than the fast rotating ones \citep[e.g.][]{2015MNRAS.450.3750M}. As braking only works efficiently in hard binaries that are expected to survive in the clusters for a Hubble time, this would imply that the young and intermediate-age clusters have fundamentally different binary properties than the Galactic globular clusters, which have binary fractions of typically a few per cent only \citep[e.g.][]{2012A&A...540A..16M}. Alternatively, a strong dependence of the properties of binaries on the masses of their constituent stars would be required, given the difference in stellar mass between the stars observed in young and intermediate-age clusters and in Galactic globulars.

Until now, the sample sizes of stars with measured $v\sin i$ values have been limited, meaning that it has not been possible to study statistically representative samples of stars in order to meaningfully compare with models that include physically motivated distributions of rotational velocities. Thanks to their large field of view and the excellent throughput, modern integral-field spectrographs such as MUSE enable us to significantly increase the spectroscopic sample sizes in star clusters compared to traditional (multi-object) spectrographs \citep[e.g.][]{2018MNRAS.473.5591K}. In an initial study presented in \citet{2018MNRAS.480.1689K}, we used adaptive optics enhanced VLT/MUSE observations of the intermediate age ($\sim1.5$~~Gyr) cluster NGC~419 in the Small Magellanic Cloud (SMC).  We found a clear decrease in effective temperatures across the eMSTO, in agreement with expectations of stellar models that include rapid rotation.  Due to the somewhat low S/N of the data, we were unable to measure the $v\sin i$ values of individual stars on the eMSTO in a reliable way.  Instead, we co-added the spectra of stars on the blue and red sides of the eMSTO and carried out a differential analysis of the resulting spectra.  We found, in agreement with rotating stellar models, that the blue spectrum had a warmer effective temperature and lower rotational velocity than the red spectrum.  Here we expand this initial study to another intermediate age cluster with significantly higher S/N data, allowing us to determine the stellar parameters and $v\sin i$ values of large numbers of individual stars across the eMSTO.

In the present work, we focus on the massive cluster NGC~1846 in the Large Magellanic Cloud (LMC).  The clusters has an age of about $1.5$~Gyr \citep[e.g.][]{2008ApJ...681L..17M} and a mass of about $2 \times 10^5$~M$_{\odot}$ \citep[e.g.][]{2014ApJ...797...35G}.  The cluster hosts an extended main sequence turn-off, which shows some evidence for being bi-modal \citep{2008ApJ...681L..17M}.  However, there is no sign of a dual main sequence, which is typically only seen in the CMDs of younger clusters.

This paper is organised as follows. We present our data in Sect.~\ref{sec:data} and their analysis in Sect.~\ref{sec:analysis}. The results are described in Sect.~\ref{sec:results} and discussed in Sect.~\ref{sec:discussion} before we conclude in Sect.~\ref{sec:conclusions}. In the reduced data, we further detected a previously unknown planetary nebula. This discovery is presented in Appendix~\ref{app:nebula}.

\section{Data}
\label{sec:data}

We observed NGC~1846 with the Multi-Unit Spectroscopic Explorer (MUSE), a panoramic integral-field spectrograph at the ESO Very Large Telescope \citep{2010SPIE.7735E..08B}. The observations were carried out during four nights, 2018-10-01, 2019-03-11, 2019-03-14, and 2019-08-22. Each observation consisted of three exposures of 880~s integration time each, targeting the central $1\arcmin\times1\arcmin$ of NGC~1846. To homogenize the image quality across the field of view, we applied a spatial dither pattern and derotator offsets of $90^\circ$ between the exposures. All observations were performed with the GALACSI adaptive optics system \citep{2016SPIE.9909E..2ZL}, in order to correct the ground-layer turbulence and enhance the spatial resolution of the data. The effective seeing in the SDSS \textit{r}-band was measured as $0\farcs5$, $0\farcs8$, $0\farcs8$, and $0\farcs7$, respectively, for the four nights.

\begin{figure}
	\includegraphics[width=\columnwidth]{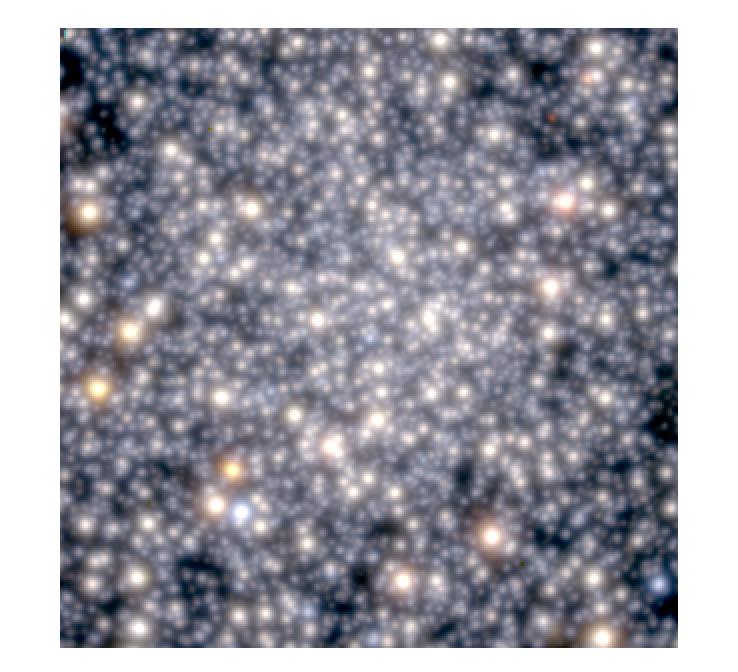}
    \caption{\textit{gri} colour image of NGC~1846 created from the MUSE data presented in this work. The image size is $1\,{\rm arcmin}^2$, North is up and East is left.}
    \label{fig:colour_image}
\end{figure}

The data reduction was performed with the standard MUSE pipeline \citep{2012SPIE.8451E..0BW, 2014ASPC..485..451W}. It performs all the basic reduction steps -- bias removal, slice tracing, wavelength calibration, and flat fielding -- for each individual integral-field unit (IFU). Afterwards, the data from all 24 IFUs are combined, flux-calibrated, and corrected for geometrical distortions. The final result is a data cube with a spatial sampling of $0.2\arcsec$, and covering the wavelength range from $480\,\rm{nm}$ to $930\,{\rm nm}$ with a spectral resolution of $\sim0.25\,{\rm nm}$ full width at half maximum (FWHM, corresponding to $R\sim1\,700-3\,500$). We created one data cube from the entire set of available exposures as well as four cubes containing only the exposures taken during a single night. The latter was done in order to search for binary stars via radial velocity variations. In Fig.~\ref{fig:colour_image}, we show a \textit{gri} colour image created from the cube with maximum depth. The effective seeing of this cube was measured as $0\farcs8$ at $550\,{\rm nm}$ and $0\farcs6$ at $850\,{\rm nm}$.

\section{Analysis}
\label{sec:analysis}

\subsection{Spectrum extraction}
\label{sec:extraction}

\begin{figure}
    \centering
    \includegraphics[width=\columnwidth]{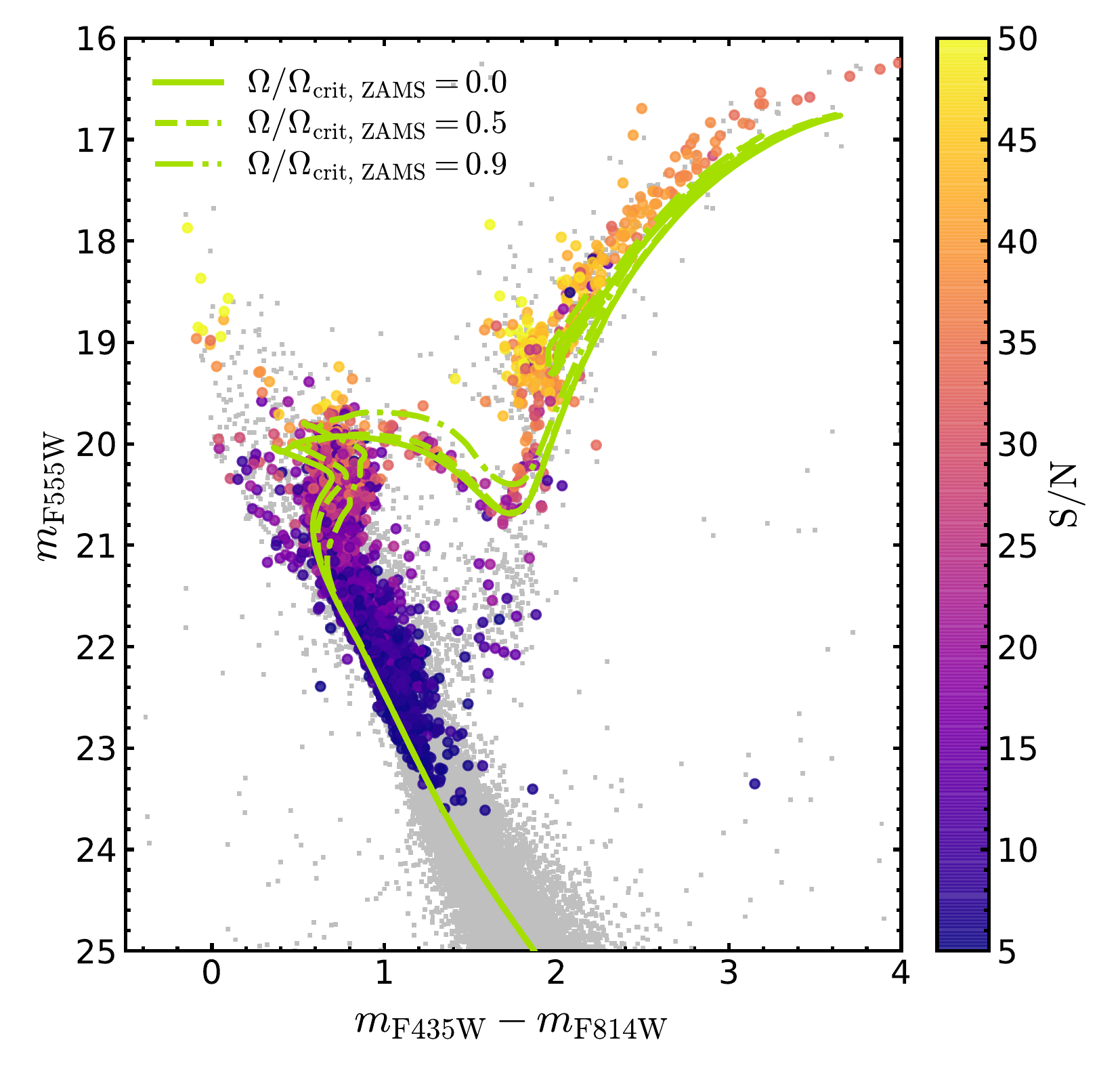}
    \caption{HST colour-magnitude diagram of NGC~1846 (grey dots), produced from the photometry presented in Sect.~\ref{sec:extraction}. The stars for which useful spectra were extracted from the MUSE data are colour-coded according to the spectral signal-to-noise per pixel, averaged over the MUSE wavelength range. The green lines show MIST isochrone predictions for an age of $\log (t/{\rm yr})=9.14$, a metallicity of $[{\rm Fe/H}]=-0.37$ ($Z=0.006$), and different stellar rotation rates $\oocrit$. See Sect.~\ref{sec:isochrone} for the definition of the critical rotation rate, $\Omega_{\rm crit,\,ZAMS}$.}
    \label{fig:cmd_extraction}
\end{figure}

We used \textsc{PampelMuse} \citep{2013A&A...549A..71K} to extract stellar spectra from the reduced MUSE cubes of NGC~1846. The code deblends the spectra of the resolved stars based on a wavelength-dependent model of the point spread function (PSF) that is recovered from the integral-field data and the source coordinates provided in an astrometric reference catalogue. For the latter, we used the \textit{Hubble Space Telescope} (HST) photometry presented in \citet{2018MNRAS.473.2688M}. It consists of Advanced Camera for Surveys (ACS) magnitudes in three filters,  F435W, F555W, and F814W. For the extraction, we relied on the sources with measured F814W magnitudes, as this passband has the largest overlap with the MUSE wavelength range.

The extraction from the data cube at full depth yielded $5\,044$ spectra. After removing spectra with low signal-to-noise (S/N < 5, measured per pixel and averaged over the MUSE wavelength range) and all stars for which the recovered F814W magnitude deviated significantly from the original one, we were left with a sample of $3\,638$ spectra that we consider for further analysis. In Fig.~\ref{fig:cmd_extraction}, we show the location of the extracted spectra in the HST colour-magnitude diagram of NGC~1846. We are able to extract spectra down to a magnitude of $m_{\rm F555W}\sim23$, approximately $2.5$ magnitudes below the eMSTO of NGC~1846. Spectra of eMSTO stars typically have ${\rm S/N}\sim25$ in the MUSE data.

We repeated the extraction process for the four cubes created from the data taken during the individual nights. Applying the same quality cuts as before, we are left with $3\,331$ spectra for night 2018-10-01, $2\,091$ spectra for night 2019-03-11, $2\,162$ spectra for night 2019-03-14, and $2\,374$ for night 2019-08-22. The different numbers can be explained by the different seeing conditions during the individual nights.

\subsection{Isochrone fitting}
\label{sec:isochrone}

The isochrones we used to fit the aforementioned HST photometry of NGC~1846 are the same as those described in \citet{2019arXiv190711251G}. Briefly, these models are an expanded version of those of MIST \citep{2016ApJ...823..102C}. The main difference is that these models contain a greater range of initial rotation rates, going from $\oocrit=0.0$ to $0.9$, in 0.1~dex steps, as well as a tailored stellar mass and metallicity range. The boundary conditions for these models are set by ATLAS12, while SYNTHE is used for the bolometric corrections \citep{1970SAOSR.309.....K,1993sssp.book.....K}. These models are evolved until the end of core helium burning.

The rotation rate $\oocrit$ is the equatorial angular velocity of the star, divided by its critical value as defined by \citet{2016ApJ...823..102C}, at the zero age main sequence (ZAMS). The critical value, $\Omega_{\rm{crit,\,ZAMS}}$, is the velocity where centrifugal forces equal the surface gravity and is an intrinsic property of the star. Rotation is initiated at the ZAMS for these models, with the specified value (e.g., $\oocrit=0.9$). The effects of gravity darkening \citep{1924MNRAS..84..665V} are included, following \citet{2011A&A...533A..43E}. Unless a viewing angle is chosen, the models assume a surface averaged effect on the luminosity and temperature of the star due to gravity darkening. No viewing angle was assumed for the comparisons carried out in this work.

Furthermore, at cluster ages around $1.5\,$Gyr, turn-off stars have low enough mass to possess surface convection layers, leading to surface magnetic fields. These magnetic fields can interact with stellar winds, magnetically braking the rotation of the star. Our models do not include an explicit treatment of magnetic braking. Instead, the rotation rate is simply reduced according to a mass-dependent factor between $1.8$ and $1.3\,{\rm M_\odot}$, going from fully to non-rotating in that mass range, respectively.  Realistic modelling of magnetic braking is being investigated \citep[e.g., see recent implementations in][]{2016A&A...587A.105A,2019A&A...622A..66G}. An implementation is planned for future generations of the MIST models.

The isochrone fits to the HST data were performed in the ($m_{\rm F435W} - m_{\rm F814W}$, $m_{\rm F555W}$) CMD shown in Fig.~\ref{fig:cmd_extraction}. We started from the values provided in \citet{2018ApJ...864L...3G} and adapted them until we achieved a good by-eye fit. Two parameter sets were found to provide a good fit. The first set is similar to the one determined by \citet{2018ApJ...864L...3G}, $Z=0.008$ ($[{\rm Fe/H}]=-0.25$) and $A_{\rm V}=0.06$, whereas the second set has lower metallicity, $Z=0.006$ ($[{\rm Fe/H}]=-0.37$), and higher extinction, $A_{\rm V}=0.26$. As cluster age, we assumed $\log({\rm age}/{\rm yr})=9.14$ in both cases. The following analyses are based on the isochrones with a metallicity of $Z=0.006$. We verified that our results do not sensitively depend on the choice of isochrone (unless noted otherwise).

We show a comparison between the HST photometry and the isochrone predictions for different $\oocrit$ ratios in Fig.~\ref{fig:cmd_extraction}. It confirms that the eMSTO of NGC~1846 can be modelled using a range of stellar rotation velocities. We observe a slight colour offset along the red giant branch (RGB) of the cluster, where the isochrone predictions are offset by $\Delta(m_{\rm F435W}-m_{\rm F814W})\sim0.1-0.2$ to the red from the data. This can be explained by an inconsistent treatment of convection in the stellar atmospheres \citep[ATLAS;][]{1970SAOSR.309.....K} and interiors \citep[MESA;][]{2011ApJS..192....3P}, which leads to a RGB effective temperature ($T_{\rm eff}$) scale that is, in general, cooler than observed. Based on the results obtained by \citet{2018ApJ...860..131C}, the next generation of MIST models will feature a self-consistent treatment of convection in MESA and ATLAS, leading to a larger convective mixing length parameter, $\alpha_{\rm MLT}$ \citep[see][section 3.6.1 for a brief description of the mixing length theory in these models]{2016ApJ...823..102C} and, thus, better agreement with observations.

We assigned each star with an available MUSE spectrum an effective temperature $T_{\rm eff, iso}$ and a surface gravity $\log g_{\rm iso}$ via comparison with the final set of isochrones. To this aim, we determined the closest isochrone point to each star in ($m_{\rm F435W} - m_{\rm F814W}$, $m_{\rm F555W}$) space and copied its values for the effective temperature and surface gravity. A small subsample of our stars are missing either the F435W or the F555W magnitude in the HST catalogue. In such cases, we assigned fixed values of $T_{\rm eff, iso}=4\,500\,{\rm K}$ and $\log g=2.5$. These values were used as initial guesses for determining the stellar parameters using full spectrum fitting (see Sect.~\ref{sec:spexxy} below).

\subsection{Cross correlation}
\label{sec:crosscorr}

We obtained initial radial velocities for the spectra extracted from each MUSE cube by cross-correlating them against synthetic PHOENIX spectra from the \textsc{G\"ottingen Spectral Library} \citep{2013A&A...553A...6H}.  For each spectrum, a matching template was selected using the stellar parameters from the isochrone comparison explained in Sect.~\ref{sec:isochrone}.

The analysis was performed with a custom \textsc{Python} programme inspired by the \textsc{fxcor} routine available in \textsc{IRAF}. Prior to the cross-correlation, the continuum was subtracted from both the object and the template spectra and wavelength regions affected by telluric absorption were set to zero. Then both spectra were resampled to the same equidistant sampling in logarithmic wavelength space. The final step before the cross-correlation was a filtering in Fourier space, removing both the lowest and highest wavenumbers.

Each cross-correlation result was assessed using the method of \citet{1979AJ.....84.1511T}. In particular, we calculated the $r^\prime$ parameter which scales with the ratio of the heights of the selected peak and an average noise peak in the asymmetric part of the cross-correlation function. As described in \citet{2014A&A...566A..58K}, $r^\prime$ can be used to estimate the uncertainties of the radial velocity measurements. There is a tendency for this method to overestimate the true uncertainties. However, the radial velocities determined via cross correlation were only used as initial guesses for the full-spectrum fitting described in Sect.~\ref{sec:spexxy} below. Hence we did not perform a calibration of their uncertainties.

We selected trustworthy results by making a simultaneous cut in $r^\prime$ and S/N, namely $r^\prime>4$ and ${\rm S/N}>5$. This resulted in a set of $2\,970$ radial velocities derived for the full-depth data set.

\subsection{Full-spectrum fitting}
\label{sec:spexxy}

\begin{figure*}
    \centering
    \includegraphics[width=\textwidth]{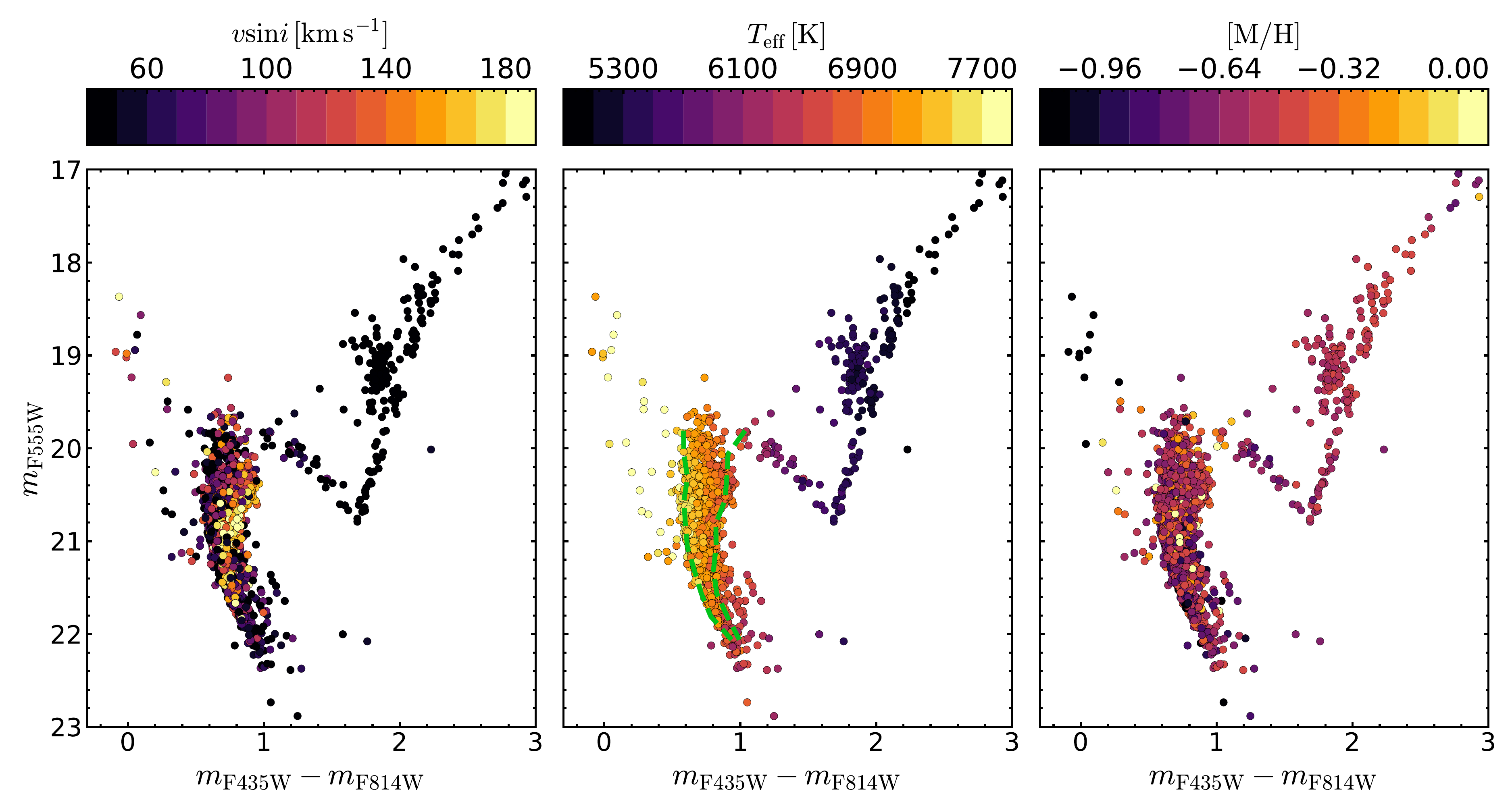}
    \caption{The distributions of projected rotational velocity (\textit{left}), effective temperature (\textit{centre}), and scaled-solar metallicity (\textit{right}), across the colour-magnitude diagram of NGC~1846. Only results derived from spectra extracted from the MUSE data with ${\rm S/N}>10$ are displayed. The results were obtained in full-spectrum fits as outlined in Sect.~\ref{sec:spexxy}. In the centre panel, dashed green lines indicate the limits used to verticalize the main sequence.}
    \label{fig:cmd_spexxy_results}
\end{figure*}

\begin{figure*}
    \centering
    \includegraphics[width=\textwidth]{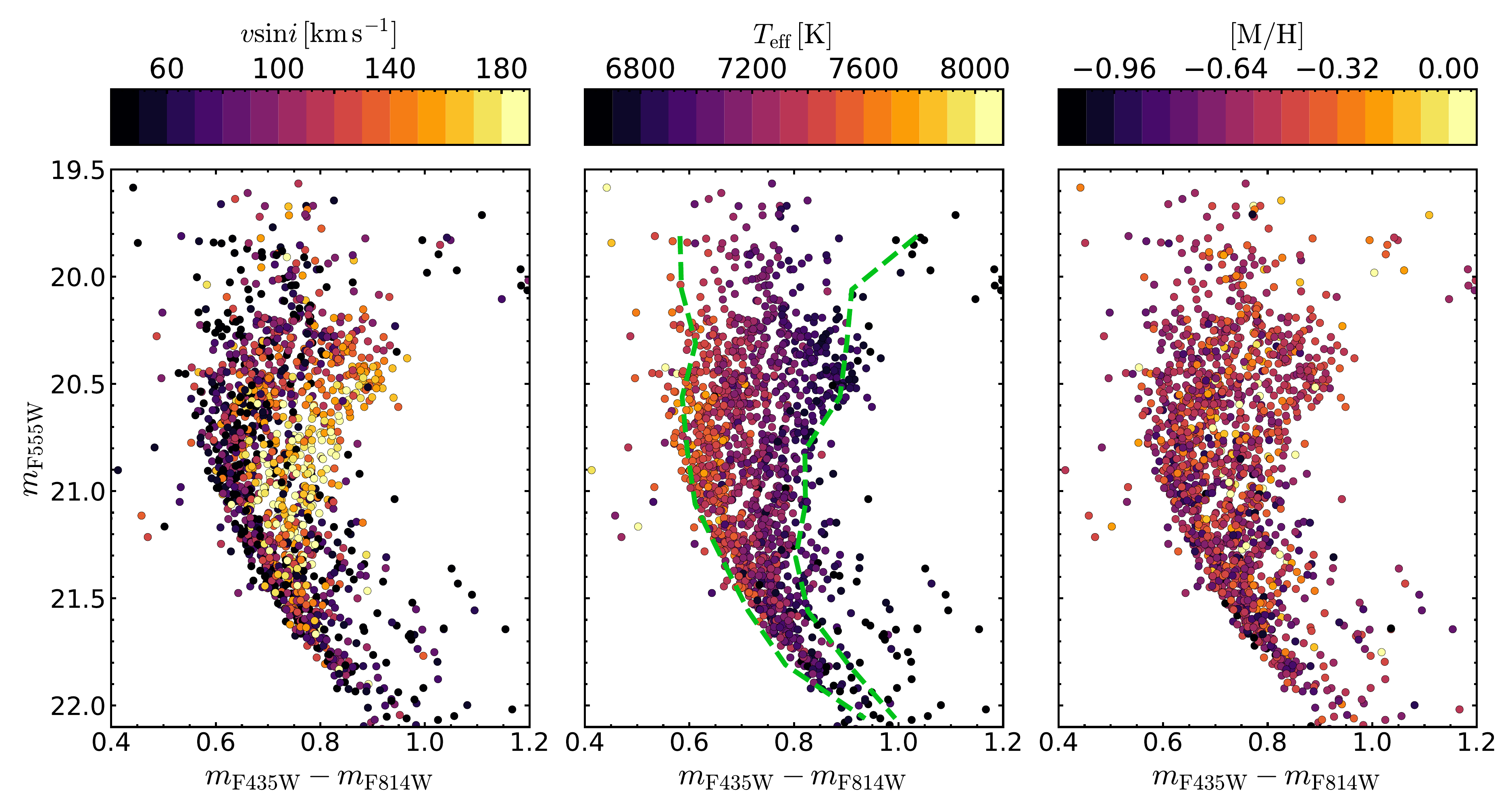}
    \caption{The same as Fig.~\ref{fig:cmd_spexxy_results}, but focused on the eMSTO region of NGC~1846.}
    \label{fig:cmd_spexxy_results_emsto_zoom}
\end{figure*}

The final step of the spectrum analysis consisted in a full-spectrum fit of each extracted spectrum. This was done using the \textsc{Spexxy} tool which is described in \citet{2016A&A...588A.148H}. It obtains a set of best-fitting stellar parameters for each spectrum by fitting it against the full \textsc{G\"ottingen Spectral Library} of PHOENIX templates. \textsc{Spexxy} comes with an interpolator that predicts the template spectrum at each point of the parameter space. It accounts for continuum mismatches between an observed spectrum and a template by a multiplicative polynomial that is applied to the template. Furthermore, \textsc{Spexxy} is also able to account for the telluric absorption bands, using a dedicated library of atmospheric spectra that are fitted simultaneously with the PHOENIX templates. For more information on the individual features, we refer to \citet{2016A&A...588A.148H}.

For each spectrum extracted from the data cube at full depth, we aimed to derive the effective temperature $T_{\rm eff}$, the solar-scaled metallicity $[{\rm M/H}]$, the radial velocity $v_{\rm los}$, and the line broadening along the line of sight, $\sigma_{\rm los}$. Recall that these quantities are derived from a comparison between the template spectra and the entire observed spectrum, not individual lines. Hence, they are tied to the strong features in the MUSE wavelength range that are most sensitive to them, such as the \ion{H}{\!$\alpha$}, \ion{H}{\!$\beta$}, and Paschen lines in case of $T_{\rm eff}$, or the \ion{Fe}{i,ii} lines bluewards of $600\,{\rm nm}$ as well as the ${\rm Mg}b$ and \ion{Ca}{ii} triplet lines in case of $[{\rm M/H}]$. All of these features also contribute to measuring $v_{\rm los}$ and $\sigma_{\rm los}$. However, owing to their narrower intrinsic widths, the discriminatory power of the metallic lines is enhanced compared to the hydrogen lines.

As initial guesses for the parameters, we used the results from the isochrone comparison described in Sect.~\ref{sec:isochrone} (for $T_{\rm eff}$ and $[{\rm M/H}]$) and from the cross correlation described in Sect.~\ref{sec:crosscorr} (for $v_{\rm los}$). In cases where the cross correlation did not yield a reliable radial velocity, we used the systemic velocity of NGC~1846 \citep[$239\,{\rm km\,s^{-1}}$,][]{2013ApJ...762...65M} as initial guess. A constant initial guess of $30\,{\rm km\,s^{-1}}$ was used for $\sigma_{\rm los}$. Because of the challenges involved in determining the surface gravity $\log g$ from low-resolution spectroscopy \citep[see][]{2016A&A...588A.148H}, we fixed $\log g$ to the values derived from the isochrone comparison during the analysis. The analysis formally converged for $3\,189$ out of the $3\,638$ useful extracted spectra.

We adopted several criteria for excluding potentially unreliable results from the converged analyses. Unless otherwise noted, we only considered results stemming from spectra extracted with ${\rm S/N}>10$. In addition, we removed stars for which the analysis yielded $\sigma_{\rm los}<5\,{\rm km\,s^{-1}}$. We found that in this regime, the uncertainties returned for the $\sigma_{\rm los}$ and $v_{\rm los}$ measurements were sometimes unreasonably high, suggesting that the optimisation got stuck at the boundary at $\sigma_{\rm los}=0$. Note that this does not imply that we can reliably measure a line broadening down to $\sigma_{\rm los}=5\,{\rm km\,s^{-1}}$. Our sensitivity limit in this respect is higher (see discussion in Sect.~\ref{sec:vsini}.

In Figs~\ref{fig:cmd_spexxy_results} and \ref{fig:cmd_spexxy_results_emsto_zoom}, we show the results obtained for $T_{\rm eff}$, $[{\rm M/H}]$, and the projected rotational velocity $v\sin i$. Only stars that are likely members of NGC~1846 (cf. Sec.~\ref{sec:membership}) and passed the aforementioned reliability cuts are shown. As mentioned earlier, instead of $v\sin i$, \textsc{Spexxy} parametrizes the line broadening via the standard deviation of the Gaussian broadening kernel, $\sigma_{\rm los}$. To convert our results, we used dedicated simulations that are detailed in Appendix~\ref{app:vsini}. We will discuss the results displayed in Figs~\ref{fig:cmd_spexxy_results} and \ref{fig:cmd_spexxy_results_emsto_zoom} in detail in Sect.~\ref{sec:results} below.

We further ran \textsc{Spexxy} on the spectra extracted from the cubes created from the observations gathered during individual nights. The main purpose of this analysis was to obtain radial velocities for the different data sets, enabling us to identify stars with variable radial velocities. For this reason, we did not try to determine $\sigma_{\rm los}$ in the single-night data sets, but fixed it to $0\,{\rm km\,s^{-1}}$. Otherwise, the analyses were performed in the same way as for the full-depth data.

To reliably separate physical radial velocity variations from those caused by the finite accuracy of our measurements, a proper knowledge of the measurement uncertainties is key. We calibrated the uncertainties from the observed epoch-to-epoch variations in a similar manner to \citet{2016A&A...588A.149K}. It is based on the idea that (in the absence of physical variations due to binary or variable stars) the normalised velocity offsets,
\begin{equation}
    \delta v_{\rm los} = \frac{v_{{\rm los},\,1}-v_{{\rm los},\,2}}{\sqrt{\epsilon_{{\rm v},\,1}^2+\epsilon_{{\rm v},\,2}^2}}\,,
\label{eq:deltav}
\end{equation}
follow a Gaussian distribution with standard deviation unity. In eq.~\ref{eq:deltav}, $v_{{\rm los},\,1},\,\epsilon_{{\rm v},\,1}$ and $v_{{\rm los},\,2},\,\epsilon_{{\rm v},\,2}$ are the velocity measurements and uncertainties obtained during two different epochs for the same star. Calibrated uncertainties are obtained by multiplying the original uncertainties with the actual standard deviation of the $\delta v_{\rm los}$ distribution.

In contrast to \citet{2016A&A...588A.149K}, where the calibration was performed by grouping stars with extracted spectra of comparable S/N, we grouped the stars according to their $T_{\rm eff}$, $\log g$, and ${\rm M/H}$ measurements. For each star, we inferred the correction factor for the formal uncertainties of its radial velocity measurements based on the distribution of the $\delta v_{\rm los}$ values of the 100 closest stars in this three-dimensional space. Variable stars were iteratively removed from the comparison sample via kappa-sigma clipping. We found the calibrated uncertainties to vary depending on the brightnesses and colours of the individual stars. Velocities of stars on the red giant branch were measured to an accuracy between $1.0\,{\rm km\,s^{-1}}$ (around $m_{\rm F555W}\sim18$) and $2.0\,{\rm km\,s^{-1}}$ (around $m_{\rm F555W}\sim20$) during each epoch. Along the main sequence, the typical uncertainties (per epoch) were larger, ranging from $5.0\,{\rm km\,s^{-1}}$ at $m_{\rm F555W}\sim20$ to $20.0\,{\rm km\,s^{-1}}$ for the faintest stars in our sample. We present the results of our variability analysis in Sect.~\ref{sec:rv_variations} below.

\subsection{Membership determination}
\label{sec:membership}

To identify non-cluster members across the $1\,{\rm arcmin}^2$ MUSE field of view, we adopted a maximum likelihood approach that takes into account the measured radial velocities as well as the surface brightness profile of NGC~1846. Briefly, we modelled the line-of-sight velocity distribution in the MUSE field of view as the superposition of two Gaussians, one representing the cluster and the other representing the field stars. In addition, each star was assigned a membership prior, corresponding to the surface density predicted by a \citet{1962AJ.....67..471K} profile at the position of the star relative to the central value of the profile. The structural parameters of the profile were taken from \citet{2009AJ....137.4988G}.

We then determined the parameters of the Gaussian profiles and a background density of stars using the affine-invariant Markov chain Monte Carlo (MCMC) sampler \texttt{emcee} \citep{2013PASP..125..306F}. Details on this process can be found in \citet{2018MNRAS.473.5591K}. With a set of optimised parameters at hand, we were able to assign each star a posterior membership probability of being a cluster member. We removed 190 stars with a probability $<0.5$ of belonging to the cluster from our sample.

When comparing the CMDs of likely members shown in Fig.~\ref{fig:cmd_spexxy_results} with the one of all extracted stars shown in Fig.~\ref{fig:cmd_extraction}, one notices that the stars along the ``bridge'' connecting the bottom of the red giant branch with the central main sequence (with $21 \lesssim m_{\rm F555W} \lesssim 23$ and $1.5\lesssim(m_{\rm F435W} - m_{\rm F814W})\lesssim 2.0$) have been discarded as non-members. Indeed, their positions in the CMD suggest that they are red giant stars from the LMC field. Note that the position of a star in the CMD was not considered in the membership selection process as this could remove potentially interesting stars in unusual locations of the CMD (such as blue stragglers or sub-subgiants) from our sample.

A detailed description of the membership determination process will be included in a forthcoming paper on the internal kinematics of NGC~1846.

\section{Results}
\label{sec:results}

\subsection{Stellar rotational velocities}
\label{sec:vsini}

\begin{figure}
    \centering
    \includegraphics[width=\columnwidth]{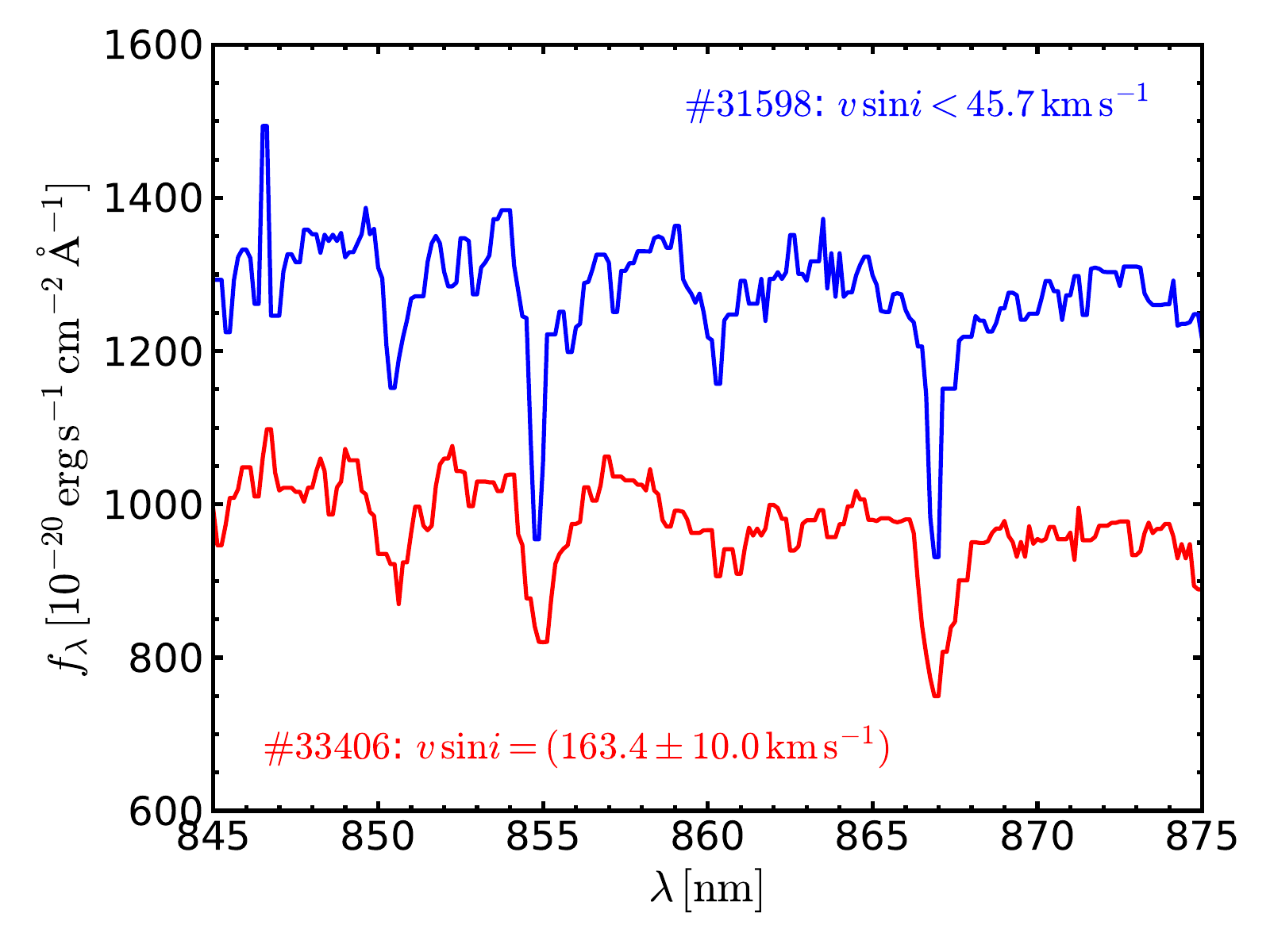}
    \caption{Comparison of the \ion{Ca}{ii} triplet region in the MUSE spectra of two eMSTO stars with very different measured rotational velocities, as indicated in the plot. Note that the spectra have been smoothed with a median filter of 3~pixels width for clarity.}
    \label{fig:msto_spectrum_comparison}
\end{figure}

\begin{figure}
    \centering
    \includegraphics[width=\columnwidth]{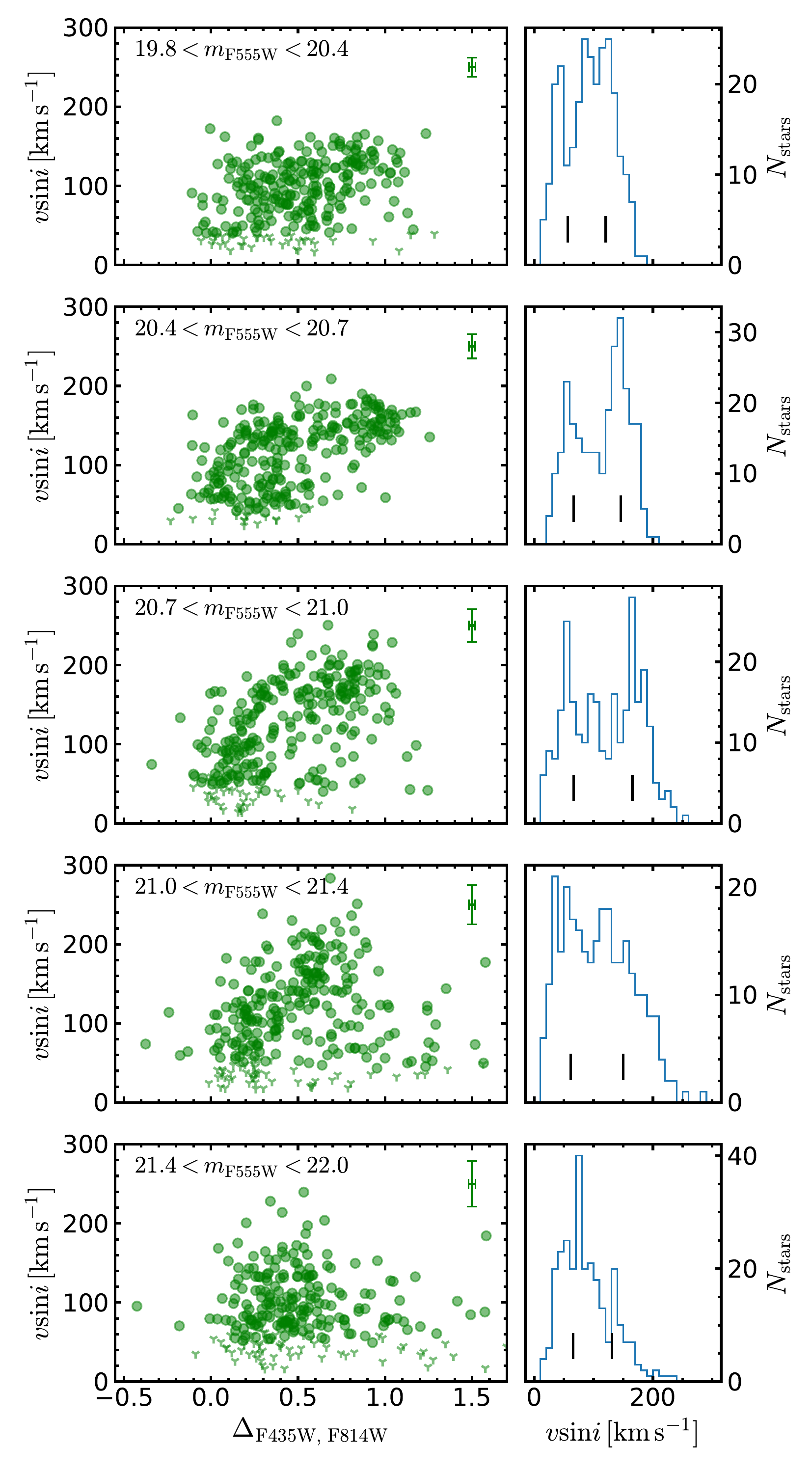}
    \caption{The distribution of projected rotational velocities, $v\sin i$, along the main sequence of NGC~1846. The left panels show the measured $v\sin i$ values as a function of the pseudo-colour $\Delta_{\rm F435W,\,F814W}$ (as defined in Sect.~\ref{sec:vsini}) for different magnitude bins as green circles, while values considered upper limits are shown as green Y-shaped symbols. Note that in order to visualise the densities of the distributions, all symbols are semi-transparent. The median uncertainty of the $v\sin i$ measurements is indicated at the top right of each panel. In the right panels, we show histograms of the $v\sin i$ measurements for the same magnitude bins as in the left panels. Black vertical lines indicate the locations of the peaks of Gaussian mixture models with $n=2$ components.
    }
    \label{fig:vsini_distributions}
\end{figure}

\begin{table*}
 \caption{Parameters of the Gaussian mixture models introduced in Sect.~\ref{sec:vsini} to represent the $v\sin i$ histograms depicted in the right panel of Fig.~\ref{fig:vsini_distributions}. The first two columns contain the lower and upper limits of the considered magnitude bins. The third column provides the average uncertainties of our $v\sin i$ measurements in each bin. Columns 4 and 6 contain the mean and standard deviation of the Gaussian component representing the slowly rotating stars, while the same properties for the component representing the fast rotating stars are listed in columns 5 and 7. Column 8 provides the fraction of stars in the fast rotating component.}
 \label{tab:vsini_distributions}
  \begin{tabular}{cccccccc}
\hline
$m_{\rm F555W,\,min}$ & $m_{\rm F555W,\,max}$ & $\overline{\epsilon_{\rm v\sin i}}$ & $\mu_{\rm slow}$ & $\mu_{\rm fast}$ & $\sigma_{\rm slow}$ & $\sigma_{\rm fast}$ & $f_{\rm fast}$\\
 &  & ${\rm km\,s^{-1}}$ & ${\rm km\,s^{-1}}$ & ${\rm km\,s^{-1}}$ & ${\rm km\,s^{-1}}$ & ${\rm km\,s^{-1}}$ & \\
\hline
19.8 & 20.4 & 12.1 & $58.3 \pm 8.2$ & $122.0 \pm 5.2$ & $24.4 \pm 13.7$ & $23.4 \pm 11.0$ &$0.55 \pm 0.09$ \\
20.4 & 20.7 & 15.3 & $66.2 \pm 3.3$ & $145.3 \pm 2.3$ & $22.3 \pm 9.1$ & $21.4 \pm 8.4$ &$0.58 \pm 0.04$ \\
20.7 & 21.0 & 20.9 & $67.0 \pm 4.2$ & $165.6 \pm 3.8$ & $29.0 \pm 11.4$ & $29.3 \pm 12.3$ &$0.51 \pm 0.04$ \\
21.0 & 21.4 & 24.9 & $61.8 \pm 9.5$ & $150.9 \pm 9.8$ & $28.0 \pm 17.6$ & $41.4 \pm 18.0$ &$0.53 \pm 0.10$ \\
21.4 & 22.0 & 28.8 & $66.1 \pm 3.0$ & $131.1 \pm 6.1$ & $24.5 \pm 8.6$ & $36.1 \pm 16.4$ &$0.37 \pm 0.05$ \\
\hline
\end{tabular}

\end{table*}

In the left panel of Fig.~\ref{fig:cmd_spexxy_results_emsto_zoom}, it is already visible that along the main sequence and in particular around the eMSTO, the stars possess a large range of rotational velocities. We measure minimum values of $v\sin i$ that are consistent with no rotation (i.e. $\lesssim40\,{\rm km\,s^{-1}}$, see below and Appendix~\ref{app:vsini}) and maximum values of $\sim250\,{\rm km\,s^{-1}}$. To illustrate how rotation alters the observed spectra, we compare in Fig.~\ref{fig:msto_spectrum_comparison} the spectra of two eMSTO stars that were fitted with low and high $v\sin i$ values, respectively. Both stars have magnitudes of $m_{\rm F555W}=20.5$, but different colours, $m_{\rm F435W} - m_{\rm F814W}=0.75$ for the slowly rotating star $\#31598$ and $m_{\rm F435W} - m_{\rm F814W}=0.90$ for the fast rotating star $\#33406$. In the spectral analysis, we found consistent metallicities and a temperature offset of $350\,{\rm K}$, with the fast rotating star being cooler, which is in line with the correlation between photometric colour and effective temperature discussed in Sect.~\ref{sec:teff} below. A significant broadening of the strong \ion{Ca}{ii} triplet lines in the spectrum of the fast rotating star is clearly visible in Fig.~\ref{fig:msto_spectrum_comparison}. This shows that despite the moderate spectral resolution of MUSE ($R\sim3\,000$ at the wavelength of the \ion{Ca}{ii} triplet, corresponding to an intrinsic FWHM of $100\,{\rm km\,s^{-1}}$), spectral differences caused by stellar rotation are detectable. However, as the simulations described in Appendix~\ref{app:vsini} show, our measurements are limited to the regime $v\sin i\gtrsim40\,{\rm km\,s^{-1}}$. For stars with projected rotational velocities below this threshold, we can only give upper limits (i.e. confirm that they are slow rotators), whereas the exact values become sensitive to systematic errors.

Fig.~\ref{fig:cmd_spexxy_results_emsto_zoom} further shows that the average $v\sin i$ increases towards the red end of the eMSTO. This is in agreement with previous studies of clusters with similar ages to NGC~1846 that found a correlation between the position of a star at the eMSTO and its $v\sin i$ \citep[e.g.][]{2018MNRAS.480.3739B,2018MNRAS.480.1689K}. A comparison to the isochrone tracks depicted in Fig.~\ref{fig:cmd_extraction} also suggests a good qualitative agreement between our measurements and the model predictions, with both showing an increase in the average $v\sin i$ from the blue to the red end of the eMSTO.

Thanks to the deep MUSE observations, we are not only able to investigate the behaviour of $v\sin i$ for eMSTO stars, but also for main-sequence stars below the turn-off. In order to investigate in-depth the behaviour of $v\sin i$ along the main sequence and its turn-off, we followed \citet{2018MNRAS.480.3739B} and determined the blue and red edges of the upper main sequence. This was done by obtaining the 5th and 95th percentiles of the distribution in $m_{\rm F435W} - m_{\rm F814W}$ colour as a function of $m_{\rm F555W}$ magnitude for the interval $19.8<m_{\rm F555W}<22.0$. Outliers, such as blue stragglers or field stars, were removed by restricting the colour distributions to $0.5<m_{\rm F435W} - m_{\rm F814W} < 1.1$.\footnote{We changed the upper limit to the 85th percentile for $m_{\rm F555W} < 21$ given the relatively large number of stars to the red of the main sequence in this magnitude range.} The resulting edges are shown in the central panels of Figs~\ref{fig:cmd_spexxy_results} and \ref{fig:cmd_spexxy_results_emsto_zoom} as dashed green lines. Afterwards, we assigned each star on the main sequence a pseudo-colour $\Delta_{\rm F435W,\,F814W}$ by measuring its normalised horizontal distance to the blue edge. We further identified as eMSTO stars all stars that passed the aforementioned selection criteria and fulfilled the condition $m_{\rm F555W}<20.6$. This resulted in a sample size of $1\,411$ main-sequence stars, out of which $474$ are eMSTO stars.

In Fig.~\ref{fig:vsini_distributions}, we analyse the variation of $v\sin i$ as a function of the $\Delta_{\rm F435W,\,F814W}$ pseudo-colour. We split up the stars in between the two fiducial lines shown in the central panels of Fig.~\ref{fig:cmd_spexxy_results} and \ref{fig:cmd_spexxy_results_emsto_zoom} into five magnitude bins, which are depicted in order of increasing magnitude in Fig.~\ref{fig:vsini_distributions}. The results for the brighter two bins, which mainly contain eMSTO stars, confirm a correlation between the $v\sin i$ of a star and its (pseudo-) colour, in the sense that stars spin faster, on average, the redder they are on the eMSTO. This trend continues down the main sequence. Only for the faintest bin shown in Fig.~\ref{fig:vsini_distributions}, no correlation between colour and projected rotational velocity is obvious. However, this bin also seems to contain significantly less fast rotating stars compared to the other bins. We suggest that this lack of fast rotators indicates that our observations have reached a regime in stellar mass where stars are efficiently spun down by magnetic braking (see discussion in Sect.~\ref{sec:discussion:vsini_dist}).

Another observation to be made in Fig.~\ref{fig:vsini_distributions} is the emergence of a separate branch of red ($\Delta_{\rm F435W,\,F814W} > \sim0.3$) but slowly rotating ($v\sin i<100\,{\rm km\,s^{-1}}$) stars. This branch appears particularly prominent in the faintest three bins, i.e. at magnitudes $m_{\rm F555W} > 20.7$. As we will discuss in Sect.~\ref{sec:discussion:binaries} below, this sequence of stars is likely to be related to binary stars.

Looking at the histograms of the $v\sin i$ measurements that are shown in the right panels of Fig.~\ref{fig:vsini_distributions}, a bimodal distribution of projected rotational velocities is suggested at least in the brighter four magnitude bins of the sample, with one maximum being located at around $v\sin i = 60\,{\rm km\,s^{-1}}$ and the other one being located at around $v\sin i = 140\,{\rm km\,s^{-1}}$. To verify the bimodality, we tried to represent the $v\sin i$ measurements in each bin with Gaussian mixture models containing $n=1$ to $n=4$ components, and used the Bayesian information criterion (BIC) to select the optimal number of components $n$ for each bin. We found that the model with $n=2$ was preferred in all five magnitude bins. The locations of the peaks suggested by this model are included in the right panels of Fig.~\ref{fig:vsini_distributions} as black vertical lines. In addition, we provide the means $\mu$ and dispersions $\sigma$ of the Gaussian components in Table~\ref{tab:vsini_distributions}. The uncertainties provided in Table~\ref{tab:vsini_distributions} were determined from 1\,000 bootstrap samples created for each magnitude bin. We find that the slowly rotating stars peak at $v\sin i\sim60\,{\rm km\,s^{-1}}$, irrespective of the magnitude bin that is considered. On the other hand, some variations with magnitude are observed for the peak of the fast rotating stars. In particular, we observe a significant drop in the mean value from $\mu_{\rm fast}=165.6\pm3.8\,{\rm km\,s^{-1}}$ to $\mu_{\rm fast}=122\pm5.2\,{\rm km\,s^{-1}}$ when going from the central to the brightest magnitude bin (cf. Table~\ref{tab:vsini_distributions}).

The Gaussian mixture models further provided us with estimates of the fractions of slow and fast rotators in each magnitude bin. As can be verified from the last column of Table~\ref{tab:vsini_distributions}, we find consistent fractions of about $55\%$ of fast rotating stars. Only for the faintest bin, the fraction is significantly lower, which we attribute to the aforementioned spin down of stars at this mass range caused by magnetic braking.

Finally, we note that at first glance, we do not measure any significant rotational velocities for stars in later evolutionary stages (cf. Fig.~\ref{fig:cmd_spexxy_results}), such as sub-giants or red giant stars. This can be explained from the strong braking of the surface such stars experience while expanding after leaving the main sequence. However, as explained in Sect.~\ref{sec:discussion:spin_down} below, the large stellar sample allows us to constrain the timescale on which the stars are being braked.

\subsection{Effective temperatures}
\label{sec:teff}

\begin{figure}
    \centering
    \includegraphics[width=\columnwidth]{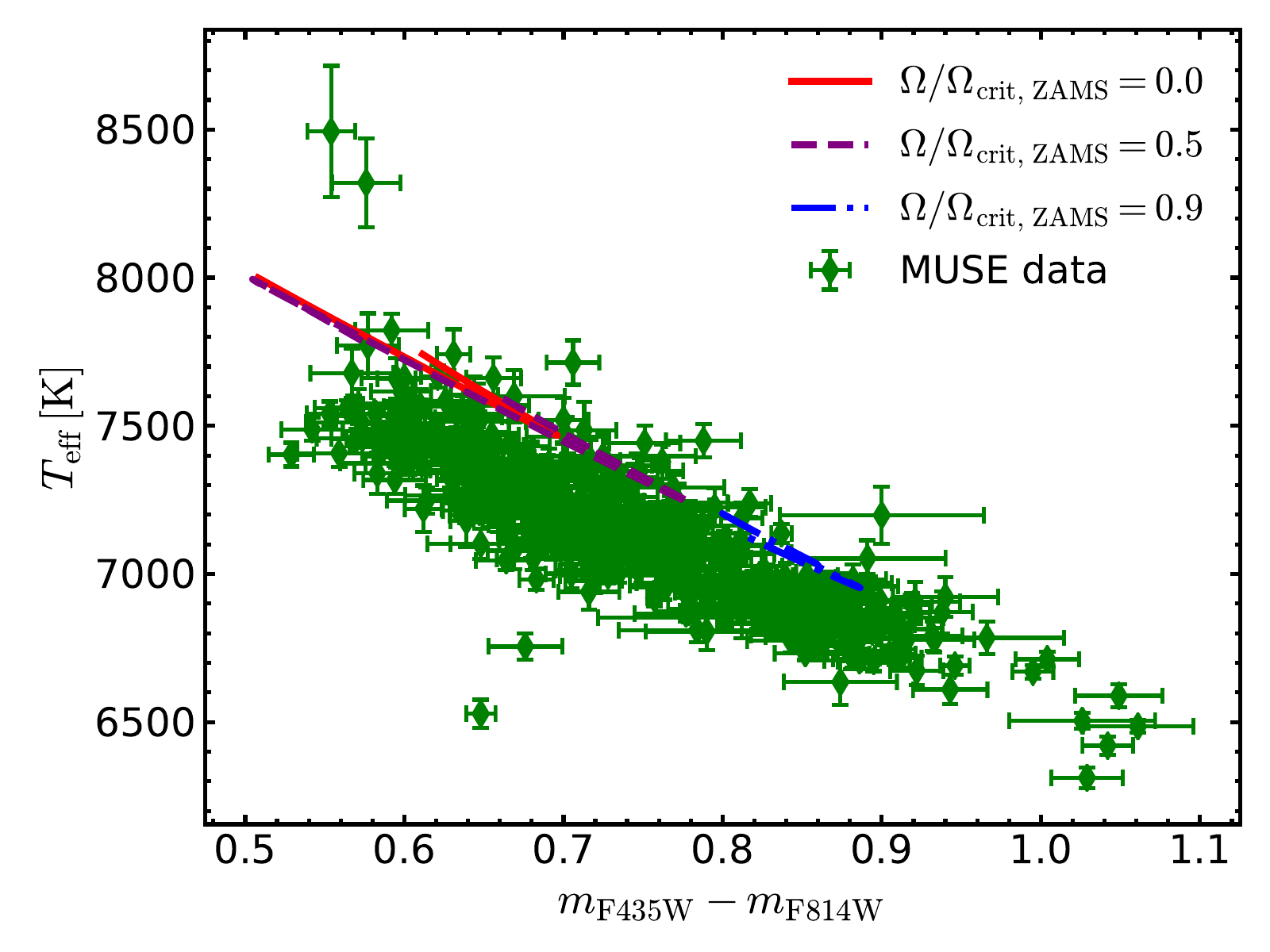}
    \caption{The distribution of effective temperatures measured across the extended main sequence turn-off of NGC~1846 as a function of photometric colour. The MUSE results are shown as green diamonds whereas the coloured lines indicate the predictions from MIST isochrones with an age of $\log({\rm age}/{\rm yr})=9.14$ and various rotation rates.}
    \label{fig:teff_emsto_distribution}
\end{figure}

We further looked at the effective temperatures measured for the stars across the eMSTO of NGC~1846. We found evidence for a temperature gradient across the eMSTO of the similar-age cluster NGC~419 in \citet{2018MNRAS.480.1689K}, as the average spectra created for the stars on both sides of the eMSTO showed a difference of $\Delta T_{\rm eff}=270\,{\rm K}$. Our analysis of the data for NGC~1846 also reveals a well-defined correlation between effective temperature and the $m_{\rm F435W}-m_{\rm F814W}$ colour, that is depicted in Fig.~\ref{fig:teff_emsto_distribution}. The end-to-end variation we observe is $\sim600\,{\rm K}$, i.e. higher than our measurement in NGC~419. However, when we split the sample in half and determine the average effective temperatures for the blue and the red subsample (similar to our analysis for NGC~419), we obtain a difference $\sim 300\,{\rm K}$, i.e comparable to what we found for NGC~419.

We also show in Fig.~\ref{fig:teff_emsto_distribution} the temperature trends predicted by the MIST isochrones presented in Sec.~\ref{sec:isochrone}, using different stellar rotation rates. While the slope of the colour-temperature relation is recovered remarkably well, the comparison to the MUSE data reveals a vertical offset of $\sim200\,{\rm K}$. The origin of this offset could be a mismatch of the selected isochrone parameters. We found that when we used isochrones with a higher metallicity ($Z=0.008$ instead of $Z=0.006$, cf. Sect.~\ref{sec:isochrone}) and lower extinction ($A_{\rm V}=0.06$ instead of $A_{\rm V}=0.26$), the predicted effective temperatures \emph{underestimated} the measured ones by $\sim200\,{\rm K}$. On the other hand, effective temperatures determined from low- or medium-resolution spectra can also be affected by systematic effects comparable to the observed offset \citep[e.g.][]{2008ApJ...682.1217K}.

\subsection{Metallicities}
\label{sec:metallicities}

The full-spectrum fits yielded a median\footnote{calculated over all likely cluster members with $S/N>10$} metallicity of $\langle[{\rm M/H}]\rangle=-0.55$ and a scatter of $\pm0.15\,{\rm dex}$, which we consider as the formal uncertainty of individual metallicity measurements. This uncertainty, however, does not account for any systematic effects in the measurements. For example, the synthetic PHOENIX spectra used in the full-spectrum fits have scaled-solar abundances. Therefore, any deviations from the assumed abundance pattern, such as an enhanced or depleted $[{\rm \alpha/Fe}]$ ratio, are likely to result in global shifts of the fitted $[{\rm M/H}]$ distribution. This could be an explanation for the difference between our median measured metallicity and the one obtained from the isochrone fits, $Z=0.006$, corresponding to $[{\rm M/H}]=-0.37$ (cf. Sect.~\ref{sec:isochrone}). For an extensive discussion of systematic effects involved in deriving metallicities from MUSE spectra, we refer to \citet{2016A&A...588A.148H}. We note that our spectroscopic measurement is in good agreement with the value of $[{\rm Fe/H}]=-0.49 \pm 0.03$ determined by \citet{2006AJ....132.1630G} using the \ion{Ca}{ii} triplet. As far as we are aware, no high-resolution measurements exist for this cluster. However, our results are within the metallicity range found by \citet{2008AJ....136..375M} for a sample of other intermediate-age LMC clusters.

We further checked if any correlation exists between $[{\rm M/H}]$ and $v\sin i$, as changes in the measured abundances with rotational velocity could help to understand how rotational mixing changes the surface structure of the stars. However, no significant correlation with $v\sin i$ was found. As mentioned above, our metallicity measurements are mainly sensitive to elements that possess strong lines in the MUSE wavelength range, such as calcium or magnesium. As it is not expected that rotational mixing has a significant impact on those elements, it is not surprising that we do not find a link between rotational velocity and surface metallicity.

\subsection{Radial velocity variations}
\label{sec:rv_variations}

\begin{figure}
    \centering
    \includegraphics[width=\columnwidth]{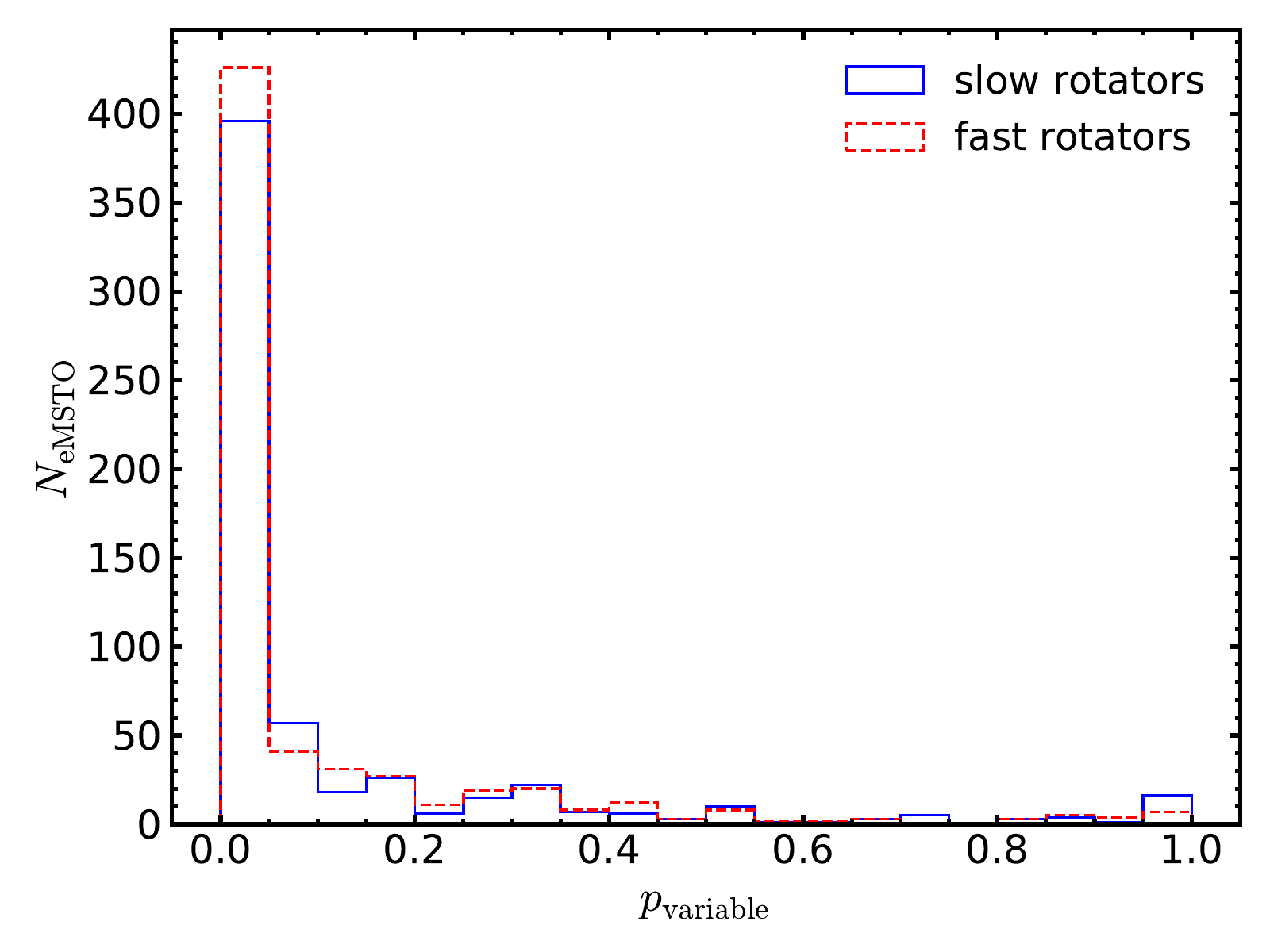}
    \caption{The probability distributions for radial-velocity variability of eMSTO stars, split according to whether the stars rotate more slowly (blue solid histogram) or faster (red dashed histrogram) than the overall sample mean.}
    \label{fig:emsto_variability}
\end{figure}

We used the radial velocities determined from the single-night spectra to see if any relation can be drawn between binarity and the measured rotational velocities. To this aim, we investigated for each main-sequence star if the night-to-night scatter in its measured radial velocity, i.e. the $\chi^2$ value obtained under the assumption of a constant velocity, was consistent with the measurement uncertainties or whether it provided evidence for intrinsic variations. This was done using the method developed by \citet{2019arXiv190904050G}, which assigns each star a probability $p$ that it is radial-velocity variable. The assignment is based upon a comparison of the observed cumulative distribution of $\chi^2$ values to the same distribution expected in the absence of binary stars. For each given $\chi^2$ value, the latter distribution gives a prediction of the number of stars expected above this value due to random variations, whereas the former distribution yields the actual number of stars observed above this value. A comparison of the two numbers then yields the probability $p$ that a star with the given $\chi^2$ is variable.

In Fig.~\ref{fig:emsto_variability}, we show the histograms of $p$ values obtained after dividing our sample of stars into two equally-sized subsamples, according to their measured $v\sin i$ relative to the median for the full sample.\footnote{We note that the results remained unchanged when we split the sample at the 45th percentile, as suggested by the fractions of fast rotators in Table~\ref{tab:vsini_distributions}.} No differences are evident between the two subsamples, a two-sided Kolmogorov-Smirnov (K-S) test yields a K-S statistic of $D=0.037$, corresponding to a probability of $p=0.80$ that the two subsamples are drawn from the same parent distribution. When we follow \citet{2019arXiv190904050G} and consider each star with $p>0.5$ as intrinsically variable, we find consistent variability fractions of $5.4\pm1.4\%$ and $7.3\pm1.5\%$ for the fast and slowly rotating subsamples, respectively.

Note that in order to convert these numbers into a binary fraction of NGC~1846, one would need to correct for the selection effects of our study \citep[see, e.g.,][]{2013A&A...550A.107S,2019arXiv190904050G}. The median uncertainties of our single-epoch velocity measurements are $13.6\,{\rm km\,s^{-1}}$ and $9.7\,{\rm km\,s^{-1}}$ for the fast and slowly rotating subsamples, respectively, and thus higher than the velocity dispersion of the cluster. On the other hand, we find a consistent binary fraction of $6.8\pm2.6\%$ among the RGB stars, for which our median single-epoch velocity measurement uncertainty is $1.9\,{\rm km\,s^{-1}}$. Nevertheless, our temporal coverage is limited, consisting of 4 epochs spread by less than a year. Hence it is likely that we missed a fraction of the present binary stars. We further note that our observations target the cluster centre, where an overdensity of binaries is expected because of mass segregation. However, to first order both populations (slow and fast rotators) should be affected in the same way by these effects. Hence it appears unlikely that the comparable binary fractions we find are a result of the aforementioned observational biases, i.e., we see no reason to doubt that the binary fractions of the two populations are most likely similar.

Binarity is not the only possible source for radial velocity variations, as they can also be caused by stellar pulsations. In fact, at the age of NGC~1846, the upper main sequence and the eMSTO crosses the instability strip, so that the presence of variable stars -- mainly $\delta$~Scuti stars -- is to be expected. \citet{2016ApJ...832L..14S} investigated the role that such stars could play in shaping the CMDs of young to intermediate-age clusters and found that a large fraction of pulsating stars would be required for them to have a detectable effect. In their study of variability in NGC~1846, \citet{2018AJ....155..183S} predicted a total number of $\sim60$ $\delta$~Scuti stars in the cluster. Unfortunately, their sample of actually detected variable stars is restricted to radii larger than the half-light radius of NGC~1846, so that there is no overlap with the MUSE sample. However, it is unlikely that we misclassify a large number of $\delta$~Scuti stars as binary candidates. $\delta$~Scuti stars typically pulsate with periods $\ll1\,{\rm d}$ and with peak-to-peak amplitudes $<10\,{\rm km\,s^{-1}}$. Such pulsation would be hard to detect with our current data. We note, however, that strongly pulsating $\delta$~Scuti stars which are most likely to affect our binary statistics are typically found to be slow rotators \citep{2000ASPC..210....3B}.

\subsection{Radial distributions}
\label{sec:radial_dist}

\begin{figure}
    \centering
    \includegraphics[width=\columnwidth]{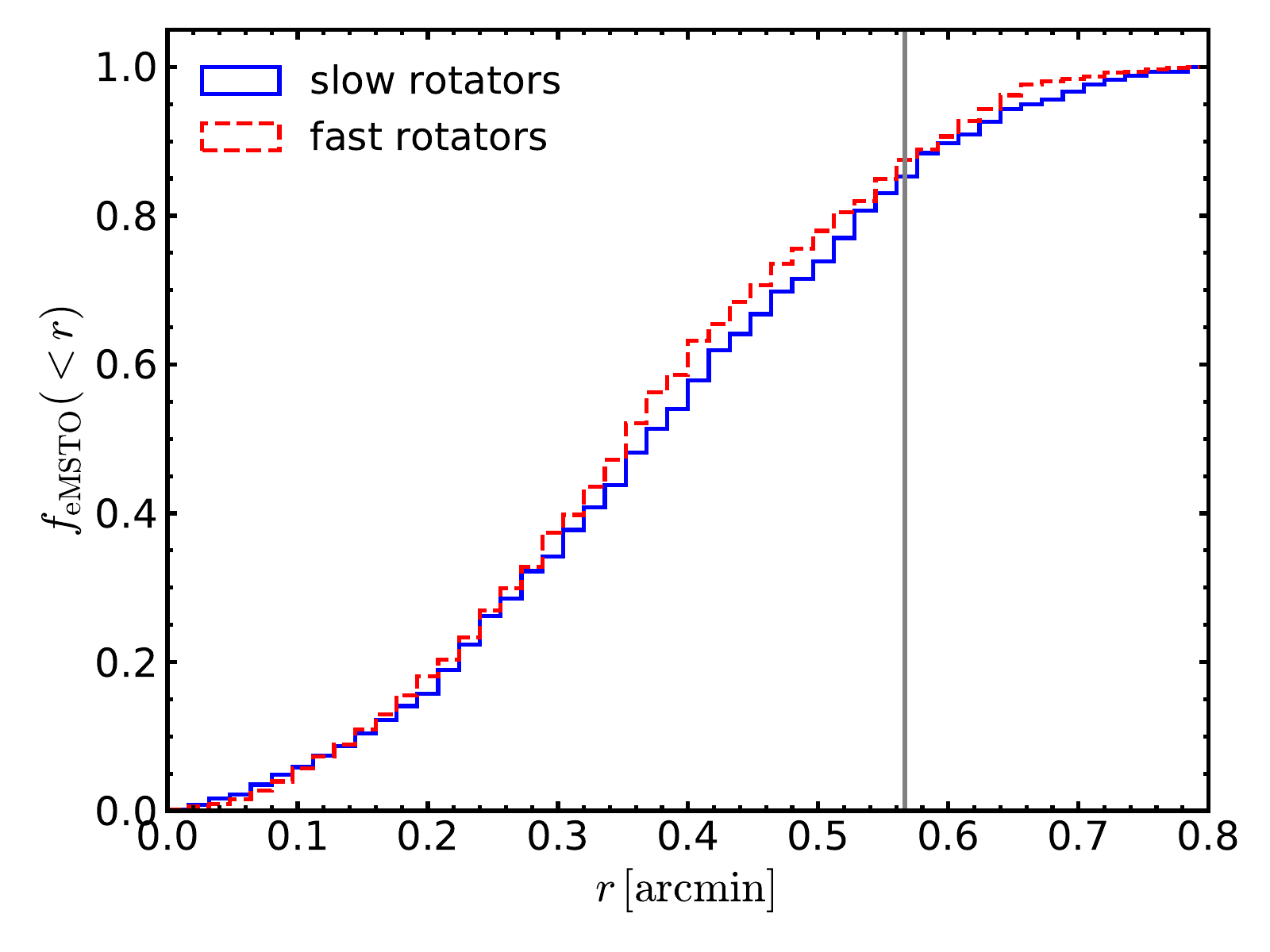}
    \caption{Cumulative fractions of slowly and fast rotating eMSTO stars in NGC~1846 as a function of distance to the cluster centre. The grey vertical line indicates the half-light radius of the cluster listed in \citet{2017MNRAS.468.3828U}.}
    \label{fig:emsto_concentrations}
\end{figure}

Finally, we looked at the radial distributions of the samples of fast and slow rotators identified in Sect.~\ref{sec:rv_variations}. NGC~1846 is among the clusters for which \citet{2011ApJ...737....4G} reported differences in the concentrations of stars across the eMSTO. This result was interpreted as evidence for an age spread among the cluster stars. However, different radial concentrations could also originate from the evolution of the cluster. For example, if any of the subpopulations had a larger binary fraction, it could appear centrally concentrated because of mass segregation.

We calculated cumulative distributions for the two subpopulations, adopting the cluster centre found by \citet{2011AJ....142...48W}. The distributions are shown in Fig.~\ref{fig:emsto_concentrations}, which illustrates that we do not detect any differences in the concentrations of slow and fast rotators. Instead, when we perform a two-sided K-S test on the distributions shown in Fig.~\ref{fig:emsto_concentrations}, we obtain a K-S statistic of $D=0.061$ and a p-value of $0.20$. Hence we cannot reject the null hypothesis that the fast and slow rotators share the same radial distributions. We note that the radial range covered by our data is a factor of $\sim2.5\times$ smaller compared to \citet{2011ApJ...737....4G} and we cannot exclude the possibility that different concentrations exist in the cluster outskirts.

\section{Discussion}
\label{sec:discussion}

\subsection{The distribution of rotational velocities}
\label{sec:discussion:vsini_dist}

\subsubsection{Rotational velocities across the eMSTO}

Our analysis in Sect.~\ref{sec:vsini} revealed a large range of projected rotational velocities amongst the eMSTO stars in NGC~1846, with values up to $v\sin i\sim250\,{\rm km\,s^{-1}}$. At the age and metallicity of NGC~1846, evolutionary models that account for stellar rotation, such as MIST (cf. Sect~\ref{sec:isochrone}) or SYCLIST \citep[e.g][]{2014A&A...566A..21G}, predict maximum surface velocities of $\sim300\,{\rm km\,s^{-1}}$ for turn-off stars that had zero-age main sequence (ZAMS) rotational velocities close to critical. The reason why our $v\sin i$ distribution does not extend to such high values is likely that $v\sin i$ measurements for stars rotating close to critical tend to underestimate the true values, because the equatorial regions, which are rotating the fastest, are gravitationally darkened \citep{2004MNRAS.350..189T}. Additionally, part of the difference could also be caused by uncertainties in the evolutionary models. Hence, we do not interpret the difference between the maximum measured and predicted rotational velocities as evidence for a lack of stars with near-critical rotation rates in NGC~1846.

The stars we observe along the eMSTO in NGC~1846 have spectral types from early F to late A. For field stars of such spectral types in the Milky Way, \citet{2007A&A...463..671R} observed distributions of $v\sin i$ extending up to $300\,{\rm km\,s^{-1}}$, albeit only very few stars were observed with $v\sin i>250\,{\rm km\,s^{-1}}$. However, a one-to-one comparison is difficult because the field star sample covers a wide range of ages and the evolution of rotation appears to be complex for stars with stellar masses around $2\,{\rm M_\odot}$ \citep{2012A&A...537A.120Z}. The gradual decline of the surface velocities for the last $\sim30\%$ of the main sequence lifetimes that was observed by \citet{2012A&A...537A.120Z} in this mass range may explain the lack of stars with $v\sin i$ values close to $300\,{\rm km\,s^{-1}}$ in our sample. This scenario is also consistent with our observations where the fast rotators in the brightest part of the eMSTO are shifted towards lower velocities  than the fainter populations (cf. Table~\ref{tab:vsini_distributions}), suggesting that they have been braked during the end of their main sequence lifetimes.

Our finding of a clear link between the position of a star at the eMSTO of NGC~1846 and its $v\sin i$ agrees with previous results in intermediate-age clusters \citep{2018MNRAS.480.3739B,2018MNRAS.480.1689K,2019ApJ...876..113S}.  It highlights that stellar rotation is indeed the dominant mechanism for creating the observed colour spreads along the upper main sequences of such clusters, confirming the scenario initially put forward by \citet{2009MNRAS.398L..11B}.

We note that the correlation between $v\sin i$ and the colour of a star is non-linear. As can be verified from the left panels of Fig.~\ref{fig:vsini_distributions}, a step increase in $v\sin i$ is observed near the blue edges of the (pseudo-)colour distributions, implying that up to $\sim 150\,{\rm km\,s^{-1}}$, the effect of $v\sin i$ on the observed colour is small. Only for higher projected rotational velocities, a stronger change in colour arises, evident from the flattening of the distributions shown in the left panels of Fig.~\ref{fig:vsini_distributions} above this value. This behaviour is in agreement with previous observations and the predictions of rotating stellar models \citep[e.g., see Fig.~5 of][]{2018MNRAS.480.3739B}.

\subsubsection{The lower mass limit of fast rotating stars}

Age spreads have been proposed as an alternative explanation for eMSTOs, although this scenario appears at odds with the observed correlation between cluster age and eMSTO width \citep{2015MNRAS.453.2070N}. Recently, \citet{2018ApJ...864L...3G} proposed that both stellar rotation and age spreads play a role in shaping the CMDs of intermediate-age clusters. The idea is based on the presence of a \emph{kink} in the main sequences of young massive clusters, which \citet{2018ApJ...864L...3G} attribute to the disappearance of rotational effects in stars fainter than the magnitude of the kink. For NGC~1846, \citet{2018ApJ...864L...3G} predict rotational effects to become negligible at $m_{\rm F555W}\sim 21$, i.e. at a magnitude where the eMSTO is already established. However, as visible in Fig.~\ref{fig:cmd_spexxy_results_emsto_zoom}, we detect fast rotating stars well below this limit, all the way down to the onset of the eMSTO at $m_{\rm F555W}\sim21.5$. Hence, we conclude that there is no need to invoke age spreads to explain the eMSTO morphology of NGC~1846.

Regardless, the presence of fast rotating stars is expected to cease below a certain magnitude, as stars of lower masses are efficiently spun down by magnetic braking. Interestingly, we find a lack of fast rotating stars at the faintest magnitudes covered by the MUSE data, as can be verified from the histograms shown in Fig.~\ref{fig:vsini_distributions} and the values listed in the last column of Table~\ref{tab:vsini_distributions}. Our faintest bin, where the lack of fast rotating stars is most obvious, includes mainly stars fainter than $m_{\rm F555W}=21.5$. This magnitude roughly coincides with the disappearance of the eMSTO, providing further evidence for the importance of stellar rotation in explaining its presence. At $m_{\rm F555W}=21.5$, the MIST isochrones predict a current stellar mass of $1.36\,M_{\rm \odot}$.

\subsubsection{Bimodal rotational distributions}

Our analysis provides strong evidence for a bimodal distribution of projected rotational velocities in NGC~1846. This finding appears in agreement with the results of \citet{2008ApJ...681L..17M}, who found photometric evidence for a bimodal eMSTO in this cluster. To our knowledge, this is the first detection of a $v\sin i$ bifurcation in an intermediate-age cluster. However, bifurcated $v\sin i$ distributions have been observed in young star clusters, e.g. amongst O- and B-stars in 30~Doradus \citep[see][]{2013A&A...560A..29R,2013A&A...550A.109D}. For stars of later spectral types, the populations of slow and fast rotators have been linked to the two branches of the split main sequences observed in their young host clusters \citep{2018AJ....156..116M,2019ApJ...883..182S}. Interestingly, these split main sequences are observed for a range in stellar mass that is comparable to the present MSTO mass of NGC~1846. This makes it likely that NGC~1846 showed a bimodal main sequence at younger ages.

The individual peaks in the $v\sin i$ distributions shown in Fig.~\ref{fig:vsini_distributions} appear remarkably narrow, given that one expects any distribution of \emph{intrinsic} rotational velocities to appear broadened by the different inclinations of rotation axes towards the line of sight. Hence, the narrowness of the peaks in the histograms of Fig.~\ref{fig:vsini_distributions} could indicate a non-isotropic distribution of inclination angles. To investigate if our data are consistent with an isotropic distribution of inclination angles, we created mock data sets, starting by assigning intrinsic rotational velocities of $60\,{\rm km\,s^{-1}}$ and $160\,{\rm km\,s^{-1}}$ to $45\%$ and $55\%$ of the stars, respectively. These numbers were chosen to resemble the parameters obtained for the brighter bins shown in Fig.~\ref{fig:vsini_distributions} (cf. Table~\ref{tab:vsini_distributions}). Then, we randomly chose an inclination for each star, assuming an isotropic distribution. Afterwards, we added noise to the sample, by randomly assigning each mock star the uncertainty of an observed star and shifting its $v\sin i$ accordingly. Finally, we used a two-component Gaussian mixture model and measured the widths of both components. This process was repeated 1\,000 times and the median widths were compared to those listed for the observed components in Table~\ref{tab:vsini_distributions}.

We found that the widths of the mock distributions were narrower or comparable to those of the observed ones. Hence, our analysis is consistent with a random distribution of inclination angles. However, in that case, the intrinsic distributions of rotational velocities must be rather narrow. For example, for the second brightest magnitude bin the widths of the observed components are $\sigma_{\rm low}=22.3\pm9.1\,{\rm km\,s^{-1}}$ and $\sigma_{\rm high}=21.4\pm8.4\,{\rm km\,s^{-1}}$ (see Table~\ref{tab:vsini_distributions}), whereas the mock components have widths of $\sigma_{\rm low,\,mock}=24.2\pm1.8{\rm km\,s^{-1}}$ and $\sigma_{\rm high}=23.0\pm2.1{\rm km\,s^{-1}}$. This yields upper limits for the widths of the intrinsic rotational velocity distributions of the slowly and fast rotating stars of $22.0\,{\rm km\,s^{-1}}$ and $21.2\,{\rm km\,s^{-1}}$, respectively. We conclude that we cannot differentiate between a highly peaked rotational distribution or a non-uniform inclination angle distribution. In principle, one could also use the isochrone model predictions to try and discriminate between the two scenarios \citep[see][]{2019MNRAS.tmp.2735D}. However, as noted in Sect.~\ref{sec:isochrone}, the isochrones used in this work are insensitive to inclination angle variations. Therefore, we leave this aspect to a future publication focusing on data model comparisons.

\subsection{The impact of binary stars}
\label{sec:discussion:binaries}

The question emerges of what causes the bimodality in the observed $v\sin i$ values. Interestingly, while \citet{2007A&A...463..671R} found bimodal $v\sin i$ distributions among more massive field stars of spectral types early A or late B \citep[see also][]{2013A&A...550A.109D}, the peak at low rotational velocities disappeared for stars with spectral types comparable to those of the eMSTO stars in NGC~1846. \citet{2015MNRAS.453.2637D} proposed that binary stars could be responsible for the emergence of the sequence of slowly rotating stars in clusters, under the assumption that all stars were born as fast rotators. Binaries were excluded from the samples of \citet{2007A&A...463..671R}, and \citet{2004ApJ...616..562A} found lower rotational velocities compared to single stars for F-, A-, and B-stars that were in binary systems. This is interpreted as evidence for losses of angular momentum due to tidal interactions.

When going to fainter magnitudes in Fig.~\ref{fig:vsini_distributions}, we note the emergence of a sequence of red stars ($\Delta_{\rm F435W,\,F814W} > 0.3$) that do not follow the general trend of an increase in $v\sin i$ towards the red edge of the eMSTO and instead have low rotational velocities ($v\sin i<100\,{\rm km\,s^{-1}}$). Their position relative to the main sequence suggests that those stars are photometric binary stars. This would also explain why the sequence disappears for the brightest bins depicted in Fig.~\ref{fig:vsini_distributions}, where the main sequence appears vertical so that the light contribution of a companion will not offset a star from it. Hence, our analysis suggests that the process of braking via tidal binary interactions also operates in star clusters.

But can binaries be responsible for all the slowly rotating stars in NGC~1846? \citet{2004ApJ...616..562A} found that the trend towards lower rotational velocities was restricted to close binaries, with orbital periods $<500$~days, suggesting that tidal effects are restricted to binaries in short orbits \citep[see also][]{2013ApJ...764..166D}. In a cluster like NGC~1846 with a central velocity dispersion $\sim5\,{\rm km\,s^{-1}}$, such system would be considered as hard binaries (i.e. binaries with a binding energy above the average kinematic energy of a cluster star) that are extremely unlikely to be destroyed in interactions with other stars. Therefore, essentially all slowly rotating stars should still be in binaries if braking via tidal interactions was the main reason for their presence. Given that we found $\sim45\%$ of the eMSTO stars to be slow rotators, this requires a fraction of hard binaries that seems too high to be consistent with the low binary fractions \citep[of typically a few per cent, e.g.,][]{2012A&A...540A..16M} observed in Galactic globular clusters of comparable mass and density as NGC~1846. We note, however, that all estimates of the binary fraction in Galactic globular clusters are based on stars with stellar masses considerably below the stars we probe in NGC~1846. Therefore, a strong stellar mass dependence of the properties of binary systems may serve as an explanation of this discrepancy. 

Our analysis of the radial velocity variations yielded binary  fraction estimates of about $6\%$ (cf. Sec.~\ref{sec:rv_variations}), comparable to estimates in Galactic globular clusters of similar masses to NGC~1846. In addition, we do not find evidence for an enhanced binary fraction among the slowly rotating stars. At first glance, this result seems at odds with the finding that photometrically detected binary stars are preferentially slow rotators. However, as discussed by \citet{2019arXiv190904050G}, the MUSE observations are not sensitive to the (approximately) equal-mass binaries that are detected via photometry. On the other hand, stars with relatively faint companions will only be detected via their radial velocity variations. Therefore, our two approaches to detect binary stars cannot be readily compared to each other. Nevertheless, we would have expected an enhanced fraction of binary candidates among the slowly rotating stars in our analysis of Sect.~\ref{sec:rv_variations} if all of them were to be in hard binary systems (even when considering that our sample of slowly rotating stars likely contains a small fraction of fast rotating stars observed at low inclination angles). So it is likely that additional ingredients are required to explain the presence of the slowly rotating subpopulation of NGC~1846, such as a broad distribution of natal rotational velocities.

We note that \citet{2013A&A...560A..29R,2015A&A...580A..92R} found qualitatively similar results, i.e. a bimodal distribution of rotational velocities, for both single and binary O-stars in the young cluster 30~Doradus. However, some differences were present between the distributions. In particular, the distribution of the binaries was lacking a tail towards high $v\sin i$ values that was present in the distribution of the single stars. This difference can be explained under the assumption that these fast rotators have been spun up in binary interactions \citep[see][]{2013ApJ...764..166D}.

\subsection{The spin-down of stars along the sub-giant branch}
\label{sec:discussion:spin_down}

\begin{figure}
    \centering
    \includegraphics[width=\columnwidth]{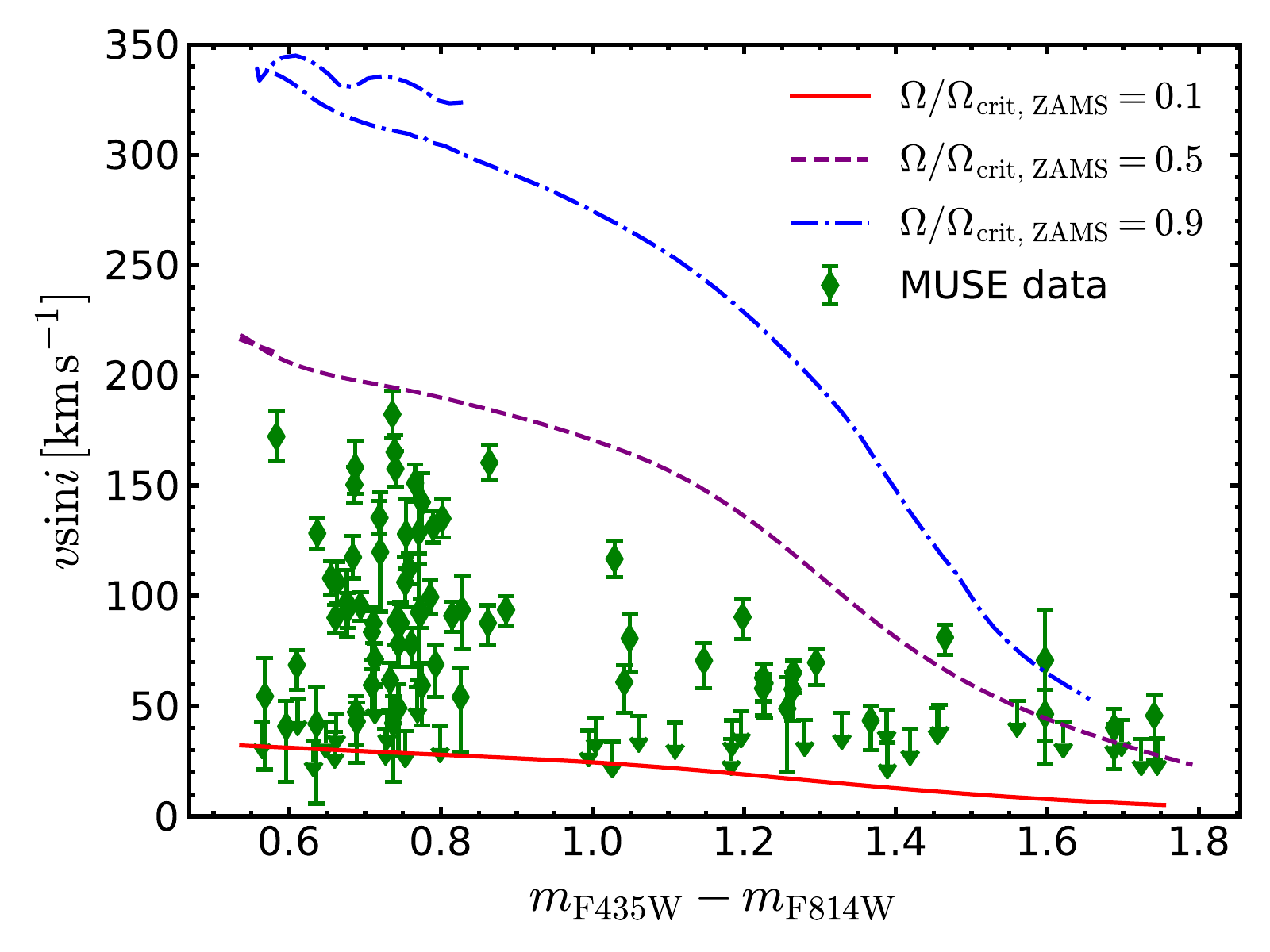}
    \caption{Projected rotational velocities of stars along the sub-giant branch of NGC~1846 as a function of photometric colour. While the MUSE measurements are shown as green diamonds, the coloured lines show the trends in average equatorial velocity predicted by the isochrones for stars with different zero-age main sequence rotation rates.}
    \label{fig:sgb_vsini_trend}
\end{figure}

As already noted in Sect.~\ref{sec:vsini}, the lack of any fast rotating stars in advanced evolutionary stages is expected, because stars leaving the main sequence are efficiently being braked during their expansion on the sub-giant branch (SGB). \citet{2016ApJ...826L..14W} previously used the morphology of the SGB to provide indirect evidence for a deceleration of fast rotating stars in the massive intermediate-age cluster NGC~419. The decrease in SGB width when moving from the blue end to the red end was interpreted as a signature of fast rotating stars being spun down. Thanks to the large MUSE sample, we are in a situation where we can study the spin-down directly, and compare the observed trend to model predictions.

To investigate how fast the stars spin down, we show in Fig.~\ref{fig:sgb_vsini_trend} the $v\sin i$ measurements obtained for sub-giants as a function of photometric colour and compare them to the trends predicted by the MIST models for different ZAMS rotation rates. The model predictions correspond to the average surface velocity of the star at the equator, and can be considered as an upper envelope (i.e. before correcting for the effects of inclination) for the measurable rotational velocity of a star with a given $\oocrit$.

At the beginning of the SGB evolution, we still observe a wide range of rotational velocities, reaching $v\sin i \sim 180\,{\rm km\,s^{-1}}$. About halfway along the SGB, the rotation rates are restricted to $v\sin i < 100\,{\rm km\,s^{-1}}$ and at the end of the SGB, no stars with $v\sin i > 50\,{\rm km\,s^{-1}}$ are observed. Hence, the spin-down appears to extend across most of the SGB evolution, which lasts about $100-200~{\rm Myr}$ for stars of this mass ($1.65\,{\rm M_\odot}$ as predicted by the MIST isochrones) and metallicity.

An interesting aspect visible in Fig.~\ref{fig:sgb_vsini_trend} is that the model prediction for a ZAMS rotation rate that is $0.5$ times the critical one appears to provide an upper limit for our data. This appears to be at odds with our finding for the eMSTO, were our $v\sin i$ measurements extended up to higher fractions of the predicted critical ZAMS rotational velocity. The reason for this difference can be seen in Fig.~\ref{fig:vsini_distributions}. In the brightest main sequence bin, the peak at high $v\sin i$ is observed at significantly lower velocities, $\mu_{\rm fast}=122\pm5\,{\rm km\,s^{-1}}$, compared to $\mu_{\rm fast}\sim150\,{\rm km\,s^{-1}}$ for the fainter bins. We interpret this behaviour as evidence that the spin-down already starts when the stars are still on the eMSTO. This also explains why in Fig.~\ref{fig:cmd_extraction} the SGB appears very narrow and no stars follow the isochrone predictions for close to critical ZAMS rotational velocities.

\section{Conclusions}
\label{sec:conclusions}

We analysed the properties of a large sample of eMSTO and main-sequence stars in the $\sim 1.5\,{\rm Gyr}$ old massive cluster NGC~1846 located in the Large Magellanic Cloud. Thanks to the large field of view of MUSE and our dedicated analysis tools, we were able to extract a large sample of $1\,411$ useful spectra of individual stars at or below the extended main sequence turn-off from the data. Compared to previous works that studied the stellar rotational velocities in young and intermediate-age clusters using 10s of stars per target, this represents an increase by a factor of $\sim 30$ in sample size. The exquisite efficiency of MUSE further allows us to study stars as faint as $m_{\rm F555W}=22$, i.e. $>2$ magnitudes below the eMSTO of NGC~1846. To our knowledge, this is the first time that genuine individual main-sequence stars in a massive intermediate-age cluster are studied spectroscopically.

Our analysis of the projected rotation velocities confirms the results from previous studies \citep[e.g.][]{2018MNRAS.480.3739B,2018MNRAS.480.1689K} that $v\sin i$ correlates with the position of a star on the eMSTO. However, we also find evidence for a bimodal rotation rate across the eMSTO, with two clearly separated peaks at $v\sin i\sim60\,{\rm km\,s^{-1}}$ and $150\,{\rm km\,s^{-1}}$. This finding appears to be in line with previous photometric studies of NGC~1846, who reported that the eMSTO also appears to be bimodal  \citep{2008ApJ...681L..17M,2009A&A...497..755M}. It is also consistent with the observation of split main sequences in young stellar clusters that are characterised by different rotation rates \citep[e.g.][]{2018AJ....156..116M}.

We investigated if the detection of two well-defined peaks in the $v\sin i$ distribution can be considered as evidence for an anisotropic distribution of inclination angles in NGC~1846, given that even for an intrinsically bimodal distribution of rotational velocities, one would expect the bimodality to be partially washed out by the different inclinations of the stars towards the line of sight. Finding evidence for non-isotropic distributions of inclination angles could provide crucial insights into cluster formation. In particular when linked to other properties like the orientation of binary systems or the rotation of a cluster as a whole, it would allow one to study how angular momentum cascades down from the scales of giant molecular clouds to stellar scales. However, our data are still consistent with an isotropic distribution of rotation angles, provided that the intrinsic distributions are considerably peaked, with widths of $\sigma\lesssim20\,{\rm km\,s^{-1}}$ (assuming a Gaussian shape).

We find evidence for the disappearance of fast rotating stars at both ends of the eMSTO. At the bright end, the spin-down of stars as they evolve along the sub-giant branch can be explained by their simultaneous expansion. Our data suggest that fast rotating eMSTO stars already start their spin-down at the end of their main sequence lives, earlier than predicted by the stellar evolutionary models. At the faint end, we observe a significant drop in the number of fast rotators for the faintest stars in our sample. We attribute this observation to the magnetic braking of stars in this magnitude range. The fact that this roughly coincides with the magnitude where the spread in photometric colour disappears suggests that stellar rotation alone can explain the morphology of the eMSTO in NGC~1846. 

The role that  binaries play in shaping the $v\sin i$ distribution of NGC~1846 remains unclear. On the one hand, we find that photometric binary stars are preferentially slow rotators, in agreement with what is found in the Galactic field. On the other hand, the total binary fraction of the slow rotators appears to be only marginally enhanced compared to the binary fraction of the fast rotators. This suggests that while stars can be spun down by tidal interactions in tight binary systems, other factors must be at play to explain the high fraction of slowly rotating stars ($\sim 45\%$) that we observe in NGC~1846.

Finally, we also report the discovery of a new planetary nebula (PN) in NGC~1846, which is most likely a cluster member (see Appendix~\ref{app:nebula}). Another PN, Mo-17 \citep[e.g.][]{2013ApJ...762...65M}, was already known in NGC~1846. Given that PNe in star clusters appear to be very rare, the coexistence of two PNe in a single cluster is quite remarkable.


\section*{Acknowledgements}

The authors are grateful for a detailed report from the anonymous referee that has led to improvements in the paper.
Based on observations made with the ESO Telescopes at the La Silla Paranal Observatory under Programme ID 0102.D-0268(A).
SK, NB, and CU gratefully acknowledge funding from a European Research Council consolidator grant (ERC-CoG-646928-Multi-Pop).
SdM acknowledges funding by the European Union's Horizon 2020 research and innovation program from the European Research Council (ERC) (Grant agreement No. 715063), and by the Netherlands Organization for Scientific Research (NWO) as part of the Vidi research program BinWaves with project number 639.042.728.
CG thanks the Equal Opportunity Office of the University of Geneva.
CL thanks the Swiss National Science Foundation for supporting this research through the Ambizione grant number PZ00P2\_168065. Support for this work was provided by NASA through Hubble Fellowship grant HST-HF2-51387.001-A awarded by the Space Telescope Science Institute, which is operated by the Association of Universities for Research in Astronomy, Inc., for NASA, under contract NAS5-26555.
This research made use of Astropy,\footnote{http://www.astropy.org} a community-developed core Python package for Astronomy \citep{2013A&A...558A..33A,2018AJ....156..123A}.
This work made use of PyAstronomy and scikit-learn \citep{scikit-learn}.




\bibliographystyle{mnras}
\bibliography{ngc1846} 

\begin{thebibliography}{}
\makeatletter
\relax
\def\mn@urlcharsother{\let\do\@makeother \do\$\do\&\do\#\do\^\do\_\do\%\do\~}
\def\mn@doi{\begingroup\mn@urlcharsother \@ifnextchar [ {\mn@doi@}
  {\mn@doi@[]}}
\def\mn@doi@[#1]#2{\def\@tempa{#1}\ifx\@tempa\@empty \href
  {http://dx.doi.org/#2} {doi:#2}\else \href {http://dx.doi.org/#2} {#1}\fi
  \endgroup}
\def\mn@eprint#1#2{\mn@eprint@#1:#2::\@nil}
\def\mn@eprint@arXiv#1{\href {http://arxiv.org/abs/#1} {{\tt arXiv:#1}}}
\def\mn@eprint@dblp#1{\href {http://dblp.uni-trier.de/rec/bibtex/#1.xml}
  {dblp:#1}}
\def\mn@eprint@#1:#2:#3:#4\@nil{\def\@tempa {#1}\def\@tempb {#2}\def\@tempc
  {#3}\ifx \@tempc \@empty \let \@tempc \@tempb \let \@tempb \@tempa \fi \ifx
  \@tempb \@empty \def\@tempb {arXiv}\fi \@ifundefined
  {mn@eprint@\@tempb}{\@tempb:\@tempc}{\expandafter \expandafter \csname
  mn@eprint@\@tempb\endcsname \expandafter{\@tempc}}}

\bibitem[\protect\citeauthoryear{{Abt} \& {Boonyarak}}{{Abt} \&
  {Boonyarak}}{2004}]{2004ApJ...616..562A}
{Abt} H.~A.,  {Boonyarak} C.,  2004, \mn@doi [\apj] {10.1086/423795}, \href
  {https://ui.adsabs.harvard.edu/abs/2004ApJ...616..562A} {616, 562}

\bibitem[\protect\citeauthoryear{{Allende Prieto} \& {del Burgo}}{{Allende
  Prieto} \& {del Burgo}}{2016}]{2016MNRAS.455.3864A}
{Allende Prieto} C.,  {del Burgo} C.,  2016, \mn@doi [\mnras]
  {10.1093/mnras/stv2518}, \href
  {https://ui.adsabs.harvard.edu/abs/2016MNRAS.455.3864A} {455, 3864}

\bibitem[\protect\citeauthoryear{{Amard}, {Palacios}, {Charbonnel}, {Gallet}
  \& {Bouvier}}{{Amard} et~al.}{2016}]{2016A&A...587A.105A}
{Amard} L.,  {Palacios} A.,  {Charbonnel} C.,  {Gallet} F.,   {Bouvier} J.,
  2016, \mn@doi [\aap] {10.1051/0004-6361/201527349}, \href
  {https://ui.adsabs.harvard.edu/abs/2016A&A...587A.105A} {587, A105}

\bibitem[\protect\citeauthoryear{{Astropy Collaboration} et~al.,}{{Astropy
  Collaboration} et~al.}{2013}]{2013A&A...558A..33A}
{Astropy Collaboration} et~al., 2013, \mn@doi [\aap]
  {10.1051/0004-6361/201322068}, \href
  {https://ui.adsabs.harvard.edu/abs/2013A%26A...558A..33A} {558, A33}

\bibitem[\protect\citeauthoryear{{Astropy Collaboration} et~al.,}{{Astropy
  Collaboration} et~al.}{2018}]{2018AJ....156..123A}
{Astropy Collaboration} et~al., 2018, \mn@doi [\aj] {10.3847/1538-3881/aabc4f},
  \href {https://ui.adsabs.harvard.edu/abs/2018AJ....156..123A} {156, 123}

\bibitem[\protect\citeauthoryear{{Bacon} et~al.,}{{Bacon}
  et~al.}{2010}]{2010SPIE.7735E..08B}
{Bacon} R.,  et~al., 2010, in Ground-based and Airborne Instrumentation for
  Astronomy III. p. 773508, \mn@doi{10.1117/12.856027}

\bibitem[\protect\citeauthoryear{{Bastian} \& {de Mink}}{{Bastian} \& {de
  Mink}}{2009}]{2009MNRAS.398L..11B}
{Bastian} N.,  {de Mink} S.~E.,  2009, \mn@doi [\mnras]
  {10.1111/j.1745-3933.2009.00696.x}, \href
  {https://ui.adsabs.harvard.edu/abs/2009MNRAS.398L..11B} {398, L11}

\bibitem[\protect\citeauthoryear{{Bastian} et~al.,}{{Bastian}
  et~al.}{2017}]{2017MNRAS.465.4795B}
{Bastian} N.,  et~al., 2017, \mn@doi [\mnras] {10.1093/mnras/stw3042}, \href
  {https://ui.adsabs.harvard.edu/abs/2017MNRAS.465.4795B} {465, 4795}

\bibitem[\protect\citeauthoryear{{Bastian}, {Kamann}, {Cabrera-Ziri}, {Georgy},
  {Ekstr{\"o}m}, {Charbonnel}, {de Juan Ovelar}  \& {Usher}}{{Bastian}
  et~al.}{2018}]{2018MNRAS.480.3739B}
{Bastian} N.,  {Kamann} S.,  {Cabrera-Ziri} I.,  {Georgy} C.,  {Ekstr{\"o}m}
  S.,  {Charbonnel} C.,  {de Juan Ovelar} M.,   {Usher} C.,  2018, \mn@doi
  [\mnras] {10.1093/mnras/sty2100}, \href
  {https://ui.adsabs.harvard.edu/abs/2018MNRAS.480.3739B} {480, 3739}

\bibitem[\protect\citeauthoryear{{Brandt} \& {Huang}}{{Brandt} \&
  {Huang}}{2015}]{2015ApJ...807...25B}
{Brandt} T.~D.,  {Huang} C.~X.,  2015, \mn@doi [\apj]
  {10.1088/0004-637X/807/1/25}, \href
  {https://ui.adsabs.harvard.edu/abs/2015ApJ...807...25B} {807, 25}

\bibitem[\protect\citeauthoryear{{Breger}}{{Breger}}{2000}]{2000ASPC..210....3B}
{Breger} M.,  2000, in {Breger} M.,  {Montgomery} M.,  eds,  Astronomical
  Society of the Pacific Conference Series Vol. 210, Delta Scuti and Related
  Stars. p.~3

\bibitem[\protect\citeauthoryear{{Chen}, {Trager}, {Peletier}, {Lan{\c c}on},
  {Vazdekis}, {Prugniel}, {Silva}  \& {Gonneau}}{{Chen}
  et~al.}{2014}]{2014A&A...565A.117C}
{Chen} Y.-P.,  {Trager} S.~C.,  {Peletier} R.~F.,  {Lan{\c c}on} A.,
  {Vazdekis} A.,  {Prugniel} P.,  {Silva} D.~R.,   {Gonneau} A.,  2014, \mn@doi
  [\aap] {10.1051/0004-6361/201322505}, \href
  {https://ui.adsabs.harvard.edu/abs/2014A%26A...565A.117C} {565, A117}

\bibitem[\protect\citeauthoryear{{Choi}, {Dotter}, {Conroy}, {Cantiello},
  {Paxton}  \& {Johnson}}{{Choi} et~al.}{2016}]{2016ApJ...823..102C}
{Choi} J.,  {Dotter} A.,  {Conroy} C.,  {Cantiello} M.,  {Paxton} B.,
  {Johnson} B.~D.,  2016, \mn@doi [\apj] {10.3847/0004-637X/823/2/102}, \href
  {https://ui.adsabs.harvard.edu/abs/2016ApJ...823..102C} {823, 102}

\bibitem[\protect\citeauthoryear{{Choi}, {Dotter}, {Conroy}  \& {Ting}}{{Choi}
  et~al.}{2018}]{2018ApJ...860..131C}
{Choi} J.,  {Dotter} A.,  {Conroy} C.,   {Ting} Y.-S.,  2018, \mn@doi [\apj]
  {10.3847/1538-4357/aac435}, \href
  {https://ui.adsabs.harvard.edu/abs/2018ApJ...860..131C} {860, 131}

\bibitem[\protect\citeauthoryear{{D'Antona}, {Di Criscienzo}, {Decressin},
  {Milone}, {Vesperini}  \& {Ventura}}{{D'Antona}
  et~al.}{2015}]{2015MNRAS.453.2637D}
{D'Antona} F.,  {Di Criscienzo} M.,  {Decressin} T.,  {Milone} A.~P.,
  {Vesperini} E.,   {Ventura} P.,  2015, \mn@doi [\mnras]
  {10.1093/mnras/stv1794}, \href
  {https://ui.adsabs.harvard.edu/abs/2015MNRAS.453.2637D} {453, 2637}

\bibitem[\protect\citeauthoryear{{Dufton} et~al.,}{{Dufton}
  et~al.}{2013}]{2013A&A...550A.109D}
{Dufton} P.~L.,  et~al., 2013, \mn@doi [\aap] {10.1051/0004-6361/201220273},
  \href {https://ui.adsabs.harvard.edu/abs/2013A&A...550A.109D} {550, A109}

\bibitem[\protect\citeauthoryear{{Dupree} et~al.,}{{Dupree}
  et~al.}{2017}]{2017ApJ...846L...1D}
{Dupree} A.~K.,  et~al., 2017, \mn@doi [\apjl] {10.3847/2041-8213/aa85dd},
  \href {https://ui.adsabs.harvard.edu/abs/2017ApJ...846L...1D} {846, L1}

\bibitem[\protect\citeauthoryear{{Espinosa Lara} \& {Rieutord}}{{Espinosa Lara}
  \& {Rieutord}}{2011}]{2011A&A...533A..43E}
{Espinosa Lara} F.,  {Rieutord} M.,  2011, \mn@doi [\aap]
  {10.1051/0004-6361/201117252}, \href
  {https://ui.adsabs.harvard.edu/abs/2011A&A...533A..43E} {533, A43}

\bibitem[\protect\citeauthoryear{{Foreman-Mackey}, {Hogg}, {Lang}  \&
  {Goodman}}{{Foreman-Mackey} et~al.}{2013}]{2013PASP..125..306F}
{Foreman-Mackey} D.,  {Hogg} D.~W.,  {Lang} D.,   {Goodman} J.,  2013, \mn@doi
  [Publications of the Astronomical Society of the Pacific] {10.1086/670067},
  \href {https://ui.adsabs.harvard.edu/abs/2013PASP..125..306F} {125, 306}

\bibitem[\protect\citeauthoryear{{Georgy}, {Granada}, {Ekstr{\"o}m}, {Meynet},
  {Anderson}, {Wyttenbach}, {Eggenberger}  \& {Maeder}}{{Georgy}
  et~al.}{2014}]{2014A&A...566A..21G}
{Georgy} C.,  {Granada} A.,  {Ekstr{\"o}m} S.,  {Meynet} G.,  {Anderson} R.~I.,
   {Wyttenbach} A.,  {Eggenberger} P.,   {Maeder} A.,  2014, \mn@doi [\aap]
  {10.1051/0004-6361/201423881}, \href
  {https://ui.adsabs.harvard.edu/abs/2014A&A...566A..21G} {566, A21}

\bibitem[\protect\citeauthoryear{{Georgy} et~al.,}{{Georgy}
  et~al.}{2019}]{2019A&A...622A..66G}
{Georgy} C.,  et~al., 2019, \mn@doi [\aap] {10.1051/0004-6361/201834505}, \href
  {https://ui.adsabs.harvard.edu/abs/2019A&A...622A..66G} {622, A66}

\bibitem[\protect\citeauthoryear{{Giesers} et~al.,}{{Giesers}
  et~al.}{2019}]{2019arXiv190904050G}
{Giesers} B.,  et~al., 2019, arXiv e-prints, \href
  {https://ui.adsabs.harvard.edu/abs/2019arXiv190904050G} {p. arXiv:1909.04050}

\bibitem[\protect\citeauthoryear{{Girardi}, {Eggenberger}  \&
  {Miglio}}{{Girardi} et~al.}{2011}]{2011MNRAS.412L.103G}
{Girardi} L.,  {Eggenberger} P.,   {Miglio} A.,  2011, \mn@doi [\mnras]
  {10.1111/j.1745-3933.2011.01013.x}, \href
  {https://ui.adsabs.harvard.edu/abs/2011MNRAS.412L.103G} {412, L103}

\bibitem[\protect\citeauthoryear{{Gossage} et~al.,}{{Gossage}
  et~al.}{2019}]{2019arXiv190711251G}
{Gossage} S.,  et~al., 2019, arXiv e-prints, \href
  {https://ui.adsabs.harvard.edu/abs/2019arXiv190711251G} {}

\bibitem[\protect\citeauthoryear{{G{\"o}ttgens} et~al.,}{{G{\"o}ttgens}
  et~al.}{2019}]{2019A&A...626A..69G}
{G{\"o}ttgens} F.,  et~al., 2019, \mn@doi [\aap] {10.1051/0004-6361/201935221},
  \href {https://ui.adsabs.harvard.edu/abs/2019A%26A...626A..69G} {626, A69}

\bibitem[\protect\citeauthoryear{{Goudfrooij}, {Puzia}, {Kozhurina-Platais}  \&
  {Chandar}}{{Goudfrooij} et~al.}{2009}]{2009AJ....137.4988G}
{Goudfrooij} P.,  {Puzia} T.~H.,  {Kozhurina-Platais} V.,   {Chandar} R.,
  2009, \mn@doi [The Astronomical Journal] {10.1088/0004-6256/137/6/4988},
  \href {https://ui.adsabs.harvard.edu/abs/2009AJ....137.4988G} {137, 4988}

\bibitem[\protect\citeauthoryear{{Goudfrooij}, {Puzia}, {Chandar}  \&
  {Kozhurina-Platais}}{{Goudfrooij} et~al.}{2011}]{2011ApJ...737....4G}
{Goudfrooij} P.,  {Puzia} T.~H.,  {Chandar} R.,   {Kozhurina-Platais} V.,
  2011, \mn@doi [\apj] {10.1088/0004-637X/737/1/4}, \href
  {https://ui.adsabs.harvard.edu/abs/2011ApJ...737....4G} {737, 4}

\bibitem[\protect\citeauthoryear{{Goudfrooij} et~al.,}{{Goudfrooij}
  et~al.}{2014}]{2014ApJ...797...35G}
{Goudfrooij} P.,  et~al., 2014, \mn@doi [\apj] {10.1088/0004-637X/797/1/35},
  \href {https://ui.adsabs.harvard.edu/abs/2014ApJ...797...35G} {797, 35}

\bibitem[\protect\citeauthoryear{{Goudfrooij}, {Girardi}, {Bellini}, {Bressan},
  {Correnti}  \& {Costa}}{{Goudfrooij} et~al.}{2018}]{2018ApJ...864L...3G}
{Goudfrooij} P.,  {Girardi} L.,  {Bellini} A.,  {Bressan} A.,  {Correnti} M.,
  {Costa} G.,  2018, \mn@doi [The Astrophysical Journal]
  {10.3847/2041-8213/aada0f}, \href
  {https://ui.adsabs.harvard.edu/abs/2018ApJ...864L...3G} {864, L3}

\bibitem[\protect\citeauthoryear{{Gray}}{{Gray}}{2008}]{2008oasp.book.....G}
{Gray} D.~F.,  2008, {The Observation and Analysis of Stellar Photospheres}

\bibitem[\protect\citeauthoryear{{Grocholski}, {Cole}, {Sarajedini}, {Geisler}
  \& {Smith}}{{Grocholski} et~al.}{2006}]{2006AJ....132.1630G}
{Grocholski} A.~J.,  {Cole} A.~A.,  {Sarajedini} A.,  {Geisler} D.,   {Smith}
  V.~V.,  2006, \mn@doi [\aj] {10.1086/507303}, \href
  {https://ui.adsabs.harvard.edu/abs/2006AJ....132.1630G} {132, 1630}

\bibitem[\protect\citeauthoryear{{Husser}, {Wende-von Berg}, {Dreizler},
  {Homeier}, {Reiners}, {Barman}  \& {Hauschildt}}{{Husser}
  et~al.}{2013}]{2013A&A...553A...6H}
{Husser} T.-O.,  {Wende-von Berg} S.,  {Dreizler} S.,  {Homeier} D.,  {Reiners}
  A.,  {Barman} T.,   {Hauschildt} P.~H.,  2013, \mn@doi [\aap]
  {10.1051/0004-6361/201219058}, \href
  {https://ui.adsabs.harvard.edu/abs/2013A%26A...553A...6H} {553, A6}

\bibitem[\protect\citeauthoryear{{Husser} et~al.,}{{Husser}
  et~al.}{2016}]{2016A&A...588A.148H}
{Husser} T.-O.,  et~al., 2016, \mn@doi [\aap] {10.1051/0004-6361/201526949},
  \href {https://ui.adsabs.harvard.edu/abs/2016A%26A...588A.148H} {588, A148}

\bibitem[\protect\citeauthoryear{{Johnston}, {Aerts}, {Pedersen}  \&
  {Bastian}}{{Johnston} et~al.}{2019}]{2019arXiv191000591J}
{Johnston} C.,  {Aerts} C.,  {Pedersen} M.~G.,   {Bastian} N.,  2019, arXiv
  e-prints, \href {https://ui.adsabs.harvard.edu/abs/2019arXiv191000591J} {p.
  arXiv:1910.00591}

\bibitem[\protect\citeauthoryear{{Kamann}, {Wisotzki}  \& {Roth}}{{Kamann}
  et~al.}{2013}]{2013A&A...549A..71K}
{Kamann} S.,  {Wisotzki} L.,   {Roth} M.~M.,  2013, \mn@doi [\aap]
  {10.1051/0004-6361/201220476}, \href
  {https://ui.adsabs.harvard.edu/abs/2013A%26A...549A..71K} {549, A71}

\bibitem[\protect\citeauthoryear{{Kamann}, {Wisotzki}, {Roth}, {Gerssen},
  {Husser}, {Sandin}  \& {Weilbacher}}{{Kamann}
  et~al.}{2014}]{2014A&A...566A..58K}
{Kamann} S.,  {Wisotzki} L.,  {Roth} M.~M.,  {Gerssen} J.,  {Husser} T.-O.,
  {Sandin} C.,   {Weilbacher} P.,  2014, \mn@doi [\aap]
  {10.1051/0004-6361/201322183}, \href
  {https://ui.adsabs.harvard.edu/abs/2014A%26A...566A..58K} {566, A58}

\bibitem[\protect\citeauthoryear{{Kamann} et~al.,}{{Kamann}
  et~al.}{2016}]{2016A&A...588A.149K}
{Kamann} S.,  et~al., 2016, \mn@doi [\aap] {10.1051/0004-6361/201527065}, \href
  {https://ui.adsabs.harvard.edu/abs/2016A%26A...588A.149K} {588, A149}

\bibitem[\protect\citeauthoryear{{Kamann} et~al.,}{{Kamann}
  et~al.}{2018a}]{2018MNRAS.473.5591K}
{Kamann} S.,  et~al., 2018a, \mn@doi [Monthly Notices of the Royal Astronomical
  Society] {10.1093/mnras/stx2719}, \href
  {https://ui.adsabs.harvard.edu/abs/2018MNRAS.473.5591K} {473, 5591}

\bibitem[\protect\citeauthoryear{{Kamann} et~al.,}{{Kamann}
  et~al.}{2018b}]{2018MNRAS.480.1689K}
{Kamann} S.,  et~al., 2018b, \mn@doi [\mnras] {10.1093/mnras/sty1958}, \href
  {https://ui.adsabs.harvard.edu/abs/2018MNRAS.480.1689K} {480, 1689}

\bibitem[\protect\citeauthoryear{{King}}{{King}}{1962}]{1962AJ.....67..471K}
{King} I.,  1962, \mn@doi [The Astronomical Journal] {10.1086/108756}, \href
  {https://ui.adsabs.harvard.edu/abs/1962AJ.....67..471K} {67, 471}

\bibitem[\protect\citeauthoryear{{Kirby}, {Guhathakurta}  \& {Sneden}}{{Kirby}
  et~al.}{2008}]{2008ApJ...682.1217K}
{Kirby} E.~N.,  {Guhathakurta} P.,   {Sneden} C.,  2008, \mn@doi [\apj]
  {10.1086/589627}, \href
  {https://ui.adsabs.harvard.edu/abs/2008ApJ...682.1217K} {682, 1217}

\bibitem[\protect\citeauthoryear{{Kurucz}}{{Kurucz}}{1970}]{1970SAOSR.309.....K}
{Kurucz} R.~L.,  1970, SAO Special Report, \href
  {https://ui.adsabs.harvard.edu/abs/1970SAOSR.309.....K} {309}

\bibitem[\protect\citeauthoryear{{Kurucz}}{{Kurucz}}{1993}]{1993sssp.book.....K}
{Kurucz} R.~L.,  1993, {SYNTHE spectrum synthesis programs and line data}

\bibitem[\protect\citeauthoryear{{La Penna} et~al.,}{{La Penna}
  et~al.}{2016}]{2016SPIE.9909E..2ZL}
{La Penna} P.,  et~al., 2016, in Adaptive Optics Systems V. p. 99092Z,
  \mn@doi{10.1117/12.2232984}

\bibitem[\protect\citeauthoryear{{Mackey} \& {Broby Nielsen}}{{Mackey} \&
  {Broby Nielsen}}{2007}]{2007MNRAS.379..151M}
{Mackey} A.~D.,  {Broby Nielsen} P.,  2007, \mn@doi [\mnras]
  {10.1111/j.1365-2966.2007.11915.x}, \href
  {https://ui.adsabs.harvard.edu/abs/2007MNRAS.379..151M} {379, 151}

\bibitem[\protect\citeauthoryear{{Mackey}, {Broby Nielsen}, {Ferguson}  \&
  {Richardson}}{{Mackey} et~al.}{2008}]{2008ApJ...681L..17M}
{Mackey} A.~D.,  {Broby Nielsen} P.,  {Ferguson} A.~M.~N.,   {Richardson}
  J.~C.,  2008, \mn@doi [\apjl] {10.1086/590343}, \href
  {https://ui.adsabs.harvard.edu/abs/2008ApJ...681L..17M} {681, L17}

\bibitem[\protect\citeauthoryear{{Mackey}, {Da Costa}, {Ferguson}  \&
  {Yong}}{{Mackey} et~al.}{2013}]{2013ApJ...762...65M}
{Mackey} A.~D.,  {Da Costa} G.~S.,  {Ferguson} A.~M.~N.,   {Yong} D.,  2013,
  \mn@doi [\apj] {10.1088/0004-637X/762/1/65}, \href
  {https://ui.adsabs.harvard.edu/abs/2013ApJ...762...65M} {762, 65}

\bibitem[\protect\citeauthoryear{{Marino}, {Przybilla}, {Milone}, {Da Costa},
  {D'Antona}, {Dotter}  \& {Dupree}}{{Marino}
  et~al.}{2018}]{2018AJ....156..116M}
{Marino} A.~F.,  {Przybilla} N.,  {Milone} A.~P.,  {Da Costa} G.,  {D'Antona}
  F.,  {Dotter} A.,   {Dupree} A.,  2018, \mn@doi [\aj]
  {10.3847/1538-3881/aad3cd}, \href
  {https://ui.adsabs.harvard.edu/abs/2018AJ....156..116M} {156, 116}

\bibitem[\protect\citeauthoryear{{Martocchia} et~al.,}{{Martocchia}
  et~al.}{2018}]{2018MNRAS.473.2688M}
{Martocchia} S.,  et~al., 2018, \mn@doi [\mnras] {10.1093/mnras/stx2556}, \href
  {https://ui.adsabs.harvard.edu/abs/2018MNRAS.473.2688M} {473, 2688}

\bibitem[\protect\citeauthoryear{{Meynet} \& {Maeder}}{{Meynet} \&
  {Maeder}}{2000}]{2000A&A...361..101M}
{Meynet} G.,  {Maeder} A.,  2000, \aap, \href
  {https://ui.adsabs.harvard.edu/abs/2000A&A...361..101M} {361, 101}

\bibitem[\protect\citeauthoryear{{Milone}, {Bedin}, {Piotto}  \&
  {Anderson}}{{Milone} et~al.}{2009}]{2009A&A...497..755M}
{Milone} A.~P.,  {Bedin} L.~R.,  {Piotto} G.,   {Anderson} J.,  2009, \mn@doi
  [\aap] {10.1051/0004-6361/200810870}, \href
  {https://ui.adsabs.harvard.edu/abs/2009A%26A...497..755M} {497, 755}

\bibitem[\protect\citeauthoryear{{Milone} et~al.,}{{Milone}
  et~al.}{2012}]{2012A&A...540A..16M}
{Milone} A.~P.,  et~al., 2012, \mn@doi [\aap] {10.1051/0004-6361/201016384},
  \href {https://ui.adsabs.harvard.edu/abs/2012A%26A...540A..16M} {540, A16}

\bibitem[\protect\citeauthoryear{{Milone} et~al.,}{{Milone}
  et~al.}{2015}]{2015MNRAS.450.3750M}
{Milone} A.~P.,  et~al., 2015, \mn@doi [\mnras] {10.1093/mnras/stv829}, \href
  {https://ui.adsabs.harvard.edu/abs/2015MNRAS.450.3750M} {450, 3750}

\bibitem[\protect\citeauthoryear{{Morgan}}{{Morgan}}{1994}]{1994A&AS..103..235M}
{Morgan} D.~H.,  1994, \aaps, \href
  {https://ui.adsabs.harvard.edu/abs/1994A&AS..103..235M} {103, 235}

\bibitem[\protect\citeauthoryear{{Mucciarelli}, {Carretta}, {Origlia}  \&
  {Ferraro}}{{Mucciarelli} et~al.}{2008}]{2008AJ....136..375M}
{Mucciarelli} A.,  {Carretta} E.,  {Origlia} L.,   {Ferraro} F.~R.,  2008,
  \mn@doi [\aj] {10.1088/0004-6256/136/1/375}, \href
  {https://ui.adsabs.harvard.edu/abs/2008AJ....136..375M} {136, 375}

\bibitem[\protect\citeauthoryear{{Niederhofer}, {Georgy}, {Bastian}  \&
  {Ekstr{\"o}m}}{{Niederhofer} et~al.}{2015}]{2015MNRAS.453.2070N}
{Niederhofer} F.,  {Georgy} C.,  {Bastian} N.,   {Ekstr{\"o}m} S.,  2015,
  \mn@doi [\mnras] {10.1093/mnras/stv1791}, \href
  {https://ui.adsabs.harvard.edu/abs/2015MNRAS.453.2070N} {453, 2070}

\bibitem[\protect\citeauthoryear{{Paxton}, {Bildsten}, {Dotter}, {Herwig},
  {Lesaffre}  \& {Timmes}}{{Paxton} et~al.}{2011}]{2011ApJS..192....3P}
{Paxton} B.,  {Bildsten} L.,  {Dotter} A.,  {Herwig} F.,  {Lesaffre} P.,
  {Timmes} F.,  2011, \mn@doi [\apjs] {10.1088/0067-0049/192/1/3}, \href
  {https://ui.adsabs.harvard.edu/abs/2011ApJS..192....3P} {192, 3}

\bibitem[\protect\citeauthoryear{Pedregosa et~al.,}{Pedregosa
  et~al.}{2011}]{scikit-learn}
Pedregosa F.,  et~al., 2011, Journal of Machine Learning Research, 12, 2825

\bibitem[\protect\citeauthoryear{{Platais} et~al.,}{{Platais}
  et~al.}{2012}]{2012ApJ...751L...8P}
{Platais} I.,  et~al., 2012, \mn@doi [\apjl] {10.1088/2041-8205/751/1/L8},
  \href {https://ui.adsabs.harvard.edu/abs/2012ApJ...751L...8P} {751, L8}

\bibitem[\protect\citeauthoryear{{Ram{\'\i}rez-Agudelo}
  et~al.,}{{Ram{\'\i}rez-Agudelo} et~al.}{2013}]{2013A&A...560A..29R}
{Ram{\'\i}rez-Agudelo} O.~H.,  et~al., 2013, \mn@doi [\aap]
  {10.1051/0004-6361/201321986}, \href
  {https://ui.adsabs.harvard.edu/abs/2013A&A...560A..29R} {560, A29}

\bibitem[\protect\citeauthoryear{{Ram{\'\i}rez-Agudelo}
  et~al.,}{{Ram{\'\i}rez-Agudelo} et~al.}{2015}]{2015A&A...580A..92R}
{Ram{\'\i}rez-Agudelo} O.~H.,  et~al., 2015, \mn@doi [\aap]
  {10.1051/0004-6361/201425424}, \href
  {https://ui.adsabs.harvard.edu/abs/2015A&A...580A..92R} {580, A92}

\bibitem[\protect\citeauthoryear{{Reid} \& {Parker}}{{Reid} \&
  {Parker}}{2006}]{2006MNRAS.373..521R}
{Reid} W.~A.,  {Parker} Q.~A.,  2006, \mn@doi [\mnras]
  {10.1111/j.1365-2966.2006.11087.x}, \href
  {https://ui.adsabs.harvard.edu/abs/2006MNRAS.373..521R} {373, 521}

\bibitem[\protect\citeauthoryear{{Royer}, {Zorec}  \& {G{\'o}mez}}{{Royer}
  et~al.}{2007}]{2007A&A...463..671R}
{Royer} F.,  {Zorec} J.,   {G{\'o}mez} A.~E.,  2007, \mn@doi [\aap]
  {10.1051/0004-6361:20065224}, \href
  {https://ui.adsabs.harvard.edu/abs/2007A%26A...463..671R} {463, 671}

\bibitem[\protect\citeauthoryear{{Salinas}, {Pajkos}, {Strader}, {Vivas}  \&
  {Contreras Ramos}}{{Salinas} et~al.}{2016}]{2016ApJ...832L..14S}
{Salinas} R.,  {Pajkos} M.~A.,  {Strader} J.,  {Vivas} A.~K.,   {Contreras
  Ramos} R.,  2016, \mn@doi [\apjl] {10.3847/2041-8205/832/1/L14}, \href
  {https://ui.adsabs.harvard.edu/abs/2016ApJ...832L..14S} {832, L14}

\bibitem[\protect\citeauthoryear{{Salinas}, {Pajkos}, {Vivas}, {Strader}  \&
  {Contreras Ramos}}{{Salinas} et~al.}{2018}]{2018AJ....155..183S}
{Salinas} R.,  {Pajkos} M.~A.,  {Vivas} A.~K.,  {Strader} J.,   {Contreras
  Ramos} R.,  2018, \mn@doi [\aj] {10.3847/1538-3881/aab551}, \href
  {https://ui.adsabs.harvard.edu/abs/2018AJ....155..183S} {155, 183}

\bibitem[\protect\citeauthoryear{{Sana} et~al.,}{{Sana}
  et~al.}{2013}]{2013A&A...550A.107S}
{Sana} H.,  et~al., 2013, \mn@doi [\aap] {10.1051/0004-6361/201219621}, \href
  {https://ui.adsabs.harvard.edu/abs/2013A&A...550A.107S} {550, A107}

\bibitem[\protect\citeauthoryear{{Sun}, {de Grijs}, {Deng}  \& {Albrow}}{{Sun}
  et~al.}{2019a}]{2019ApJ...876..113S}
{Sun} W.,  {de Grijs} R.,  {Deng} L.,   {Albrow} M.~D.,  2019a, \mn@doi [\apj]
  {10.3847/1538-4357/ab16e4}, \href
  {https://ui.adsabs.harvard.edu/abs/2019ApJ...876..113S} {876, 113}

\bibitem[\protect\citeauthoryear{{Sun}, {Li}, {Deng}  \& {de Grijs}}{{Sun}
  et~al.}{2019b}]{2019ApJ...883..182S}
{Sun} W.,  {Li} C.,  {Deng} L.,   {de Grijs} R.,  2019b, \mn@doi [\apj]
  {10.3847/1538-4357/ab3cd0}, \href
  {https://ui.adsabs.harvard.edu/abs/2019ApJ...883..182S} {883, 182}

\bibitem[\protect\citeauthoryear{{Tonry} \& {Davis}}{{Tonry} \&
  {Davis}}{1979}]{1979AJ.....84.1511T}
{Tonry} J.,  {Davis} M.,  1979, \mn@doi [\aj] {10.1086/112569}, \href
  {https://ui.adsabs.harvard.edu/abs/1979AJ.....84.1511T} {84, 1511}

\bibitem[\protect\citeauthoryear{{Townsend}, {Owocki}  \& {Howarth}}{{Townsend}
  et~al.}{2004}]{2004MNRAS.350..189T}
{Townsend} R.~H.~D.,  {Owocki} S.~P.,   {Howarth} I.~D.,  2004, \mn@doi
  [\mnras] {10.1111/j.1365-2966.2004.07627.x}, \href
  {https://ui.adsabs.harvard.edu/abs/2004MNRAS.350..189T} {350, 189}

\bibitem[\protect\citeauthoryear{{Usher} et~al.,}{{Usher}
  et~al.}{2017}]{2017MNRAS.468.3828U}
{Usher} C.,  et~al., 2017, \mn@doi [\mnras] {10.1093/mnras/stx713}, \href
  {https://ui.adsabs.harvard.edu/abs/2017MNRAS.468.3828U} {468, 3828}

\bibitem[\protect\citeauthoryear{{Weilbacher}, {Streicher}, {Urrutia}, {Jarno},
  {P{\'e}contal-Rousset}, {Bacon}  \& {B{\"o}hm}}{{Weilbacher}
  et~al.}{2012}]{2012SPIE.8451E..0BW}
{Weilbacher} P.~M.,  {Streicher} O.,  {Urrutia} T.,  {Jarno} A.,
  {P{\'e}contal-Rousset} A.,  {Bacon} R.,   {B{\"o}hm} P.,  2012, in Software
  and Cyberinfrastructure for Astronomy II. p. 84510B,
  \mn@doi{10.1117/12.925114}

\bibitem[\protect\citeauthoryear{{Weilbacher}, {Streicher}, {Urrutia},
  {P{\'e}contal-Rousset}, {Jarno}  \& {Bacon}}{{Weilbacher}
  et~al.}{2014}]{2014ASPC..485..451W}
{Weilbacher} P.~M.,  {Streicher} O.,  {Urrutia} T.,  {P{\'e}contal-Rousset} A.,
   {Jarno} A.,   {Bacon} R.,  2014, in {Manset} N.,  {Forshay} P.,  eds,
  Astronomical Society of the Pacific Conference Series Vol. 485, Astronomical
  Data Analysis Software and Systems XXIII. p.~451 (\mn@eprint {arXiv}
  {1507.00034})

\bibitem[\protect\citeauthoryear{{Werchan} \& {Zaritsky}}{{Werchan} \&
  {Zaritsky}}{2011}]{2011AJ....142...48W}
{Werchan} F.,  {Zaritsky} D.,  2011, \mn@doi [\aj]
  {10.1088/0004-6256/142/2/48}, \href
  {https://ui.adsabs.harvard.edu/abs/2011AJ....142...48W} {142, 48}

\bibitem[\protect\citeauthoryear{{Wu}, {Li}, {de Grijs}  \& {Deng}}{{Wu}
  et~al.}{2016}]{2016ApJ...826L..14W}
{Wu} X.,  {Li} C.,  {de Grijs} R.,   {Deng} L.,  2016, \mn@doi [\apjl]
  {10.3847/2041-8205/826/1/L14}, \href
  {https://ui.adsabs.harvard.edu/abs/2016ApJ...826L..14W} {826, L14}

\bibitem[\protect\citeauthoryear{{Zorec} \& {Royer}}{{Zorec} \&
  {Royer}}{2012}]{2012A&A...537A.120Z}
{Zorec} J.,  {Royer} F.,  2012, \mn@doi [\aap] {10.1051/0004-6361/201117691},
  \href {https://ui.adsabs.harvard.edu/abs/2012A%26A...537A.120Z} {537, A120}

\bibitem[\protect\citeauthoryear{{de Juan Ovelar} et~al.,}{{de Juan Ovelar}
  et~al.}{2019}]{2019MNRAS.tmp.2735D}
{de Juan Ovelar} M.,  et~al., 2019, \mn@doi [\mnras] {10.1093/mnras/stz3128},
  \href {https://ui.adsabs.harvard.edu/abs/2019MNRAS.tmp.2735D} {p.~2735}

\bibitem[\protect\citeauthoryear{{de Mink}, {Langer}, {Izzard}, {Sana}  \& {de
  Koter}}{{de Mink} et~al.}{2013}]{2013ApJ...764..166D}
{de Mink} S.~E.,  {Langer} N.,  {Izzard} R.~G.,  {Sana} H.,   {de Koter} A.,
  2013, \mn@doi [\apj] {10.1088/0004-637X/764/2/166}, \href
  {https://ui.adsabs.harvard.edu/abs/2013ApJ...764..166D} {764, 166}

\bibitem[\protect\citeauthoryear{{von Zeipel}}{{von
  Zeipel}}{1924}]{1924MNRAS..84..665V}
{von Zeipel} H.,  1924, \mn@doi [\mnras] {10.1093/mnras/84.9.665}, \href
  {https://ui.adsabs.harvard.edu/abs/1924MNRAS..84..665V} {84, 665}

\makeatother
\end{thebibliography}




\appendix

\section{Calibration of the \texorpdfstring{$\MakeLowercase{v}\sin \MakeLowercase{i}$}{vsini} scale}
\label{app:vsini}

Our approach to measuring $v\sin i$ is different from most previous studies, where the widths of individual lines in high-resolution spectra have been determined. In particular, in the full-spectrum fit performed by \textsc{Spexxy}, the line broadening is parametrized as the standard deviation $\sigma_{\rm LOS}$ of a Gaussian broadening kernel.

To verify that we can determine $v\sin i$ even at the limited spectral resolution of the MUSE data and to calibrate our $\sigma_{\rm LOS}$ measurements on a $v\sin i$ scale, we designed dedicated simulations. One obvious approach would be to take synthetic spectra, broaden them to a certain projected rotation velocity, and analyse them in the same way as the observed spectra. However, we expect that the largest uncertainty in our analysis will be systematic mismatches between the synthetic PHOENIX templates and the observed spectra, which would not be accounted for by such a simulation. Therefore, we followed a different route and took an actual observed spectrum from the \textit{X-Shooter Spectral Library} \citep[XSL,][]{2014A&A...565A.117C}. We selected the spectrum of HD~18769 for our test as it offers the best compromise in having stellar parameters comparable to those expected for MSTO stars in NGC~1846 and having a low projected rotation velocity itself. The latter was estimated to $v\sin i\sim40\,{\rm km\,s^{-1}}$ from measuring the widths of metal lines in the XSL spectrum. The stellar parameters were determined by \citet{2016MNRAS.455.3864A} as $T_{\rm eff}=8057\,{\rm K}$, $\log g=3.63$, and $[{\rm Fe/H}]=+0.15$. Hence, HD~18769 is about $+0.5\,{\rm dex}$ more metal rich than the stars of NGC~1846.

We simulated mock MUSE spectra by broadening the wavelength-dependent MUSE line-spread function (LSF) to a certain $v\sin i$ value, according to the formula of \citet{2008oasp.book.....G}\footnote{using the implementation in \url{https://github.com/sczesla/PyAstronomy}.}. The resulting kernel was used to broaden the XSL spectrum, and  the result was resampled to the wavelength range and spectral sampling of the MUSE data. Finally, noise was added to the spectrum.

Our analysis needs to account for the intrinsic rotation of HD~18769. Analysis of the XSL spectrum yields $v\sin i_{\rm HD\,18769}\approx40\,{\rm km\,s^{-1}}$. This value needs to be added in quadrature to the $v\sin i$ values we used to generate the mock spectra in order to obtain their true rotational broadening. Note that this step assumes that the rotational broadening kernels have Gaussian shape, which is not strictly true. In addition, this step implies that we cannot use the simulations to test the recovery of values in the range $v\sin i < 40\,{\rm km\,s^{-1}}$. To address these two shortcomings, we performed a second set of simulations, using synthetic spectra, which is described below.

In total, we generated $1\,000$ mock spectra. Each was randomly assigned a rotation velocity from the interval $0 < v\sin i < 200\,{\rm km\,s^{-1}}$. The MUSE LSF was recovered from the dedicated data product (LSF\_PROFILE) generated by the MUSE pipeline. The pipeline recovers the LSF for every slice of every IFU, so we randomly selected the IFU and slice number for each simulated spectrum in order to verify how robust our results are against LSF mismatches. Furthermore, we applied a random radial velocity to each spectrum (from the interval $30\,{\rm km\,s^{-1}}<v_{\rm LOS}<70\,{\rm km\,s^{-1}}$). The S/N of the final spectra was between $10$ and $100$.

When analysing the mock spectra, we selected the initial guesses for the stellar parameters randomly from the ranges around the literature values, i.e. $\log g=3.63\pm0.10$, $T_{\rm eff}=(8057\pm100)\,{\rm K}$, and $[{\rm Fe/H}]=+0.15\pm0.10$. The initial guesses for the radial velocity were also drawn randomly from an interval around the true value, where the width of the interval scaled inversely with the S/N of the simulated spectra. In the analysis with \textsc{Spexxy}, we allowed every parameter to vary, only $\log g$ was fixed to the initial guess.

\begin{figure*}
    \centering
    \includegraphics[width=\columnwidth]{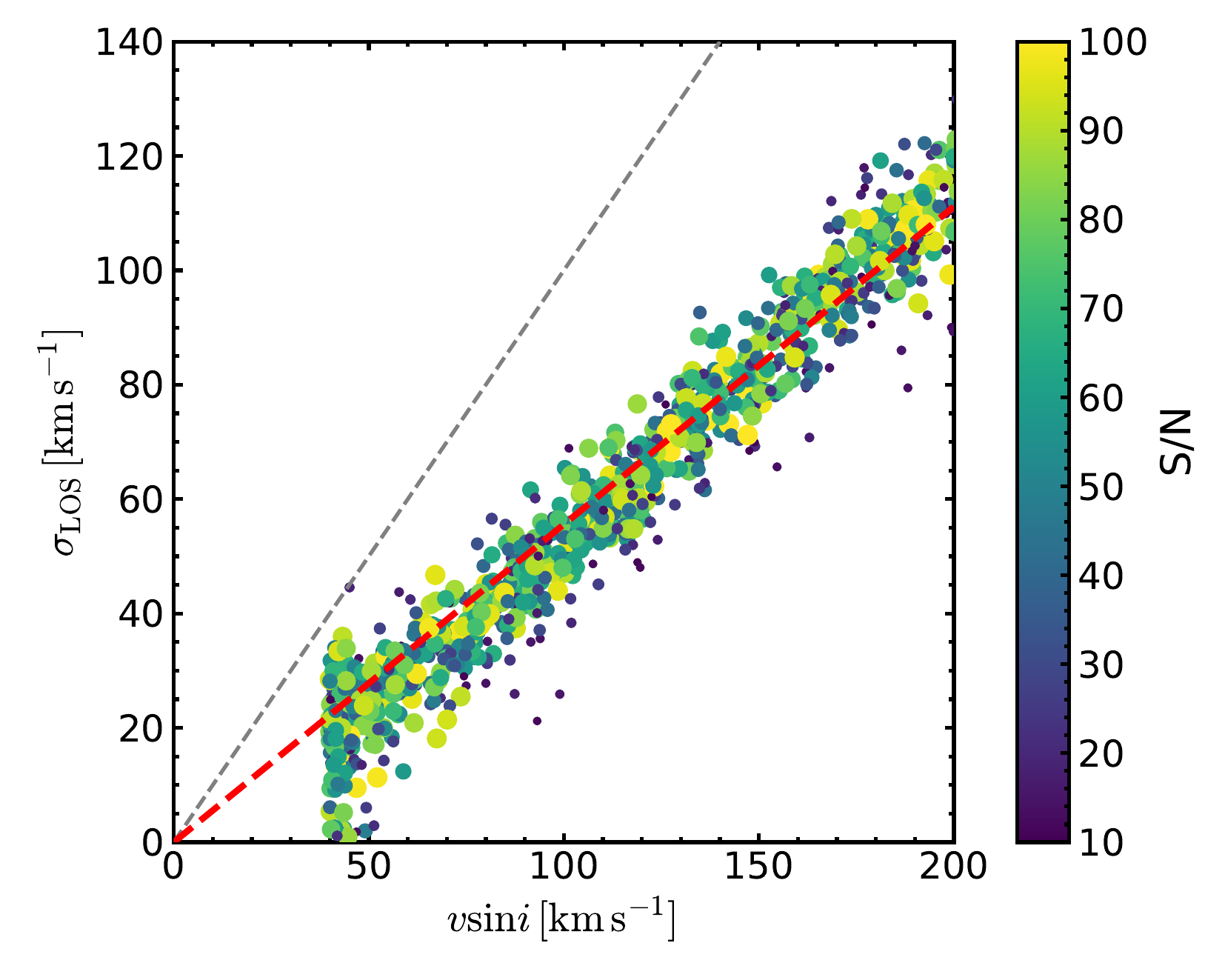}
    \includegraphics[width=\columnwidth]{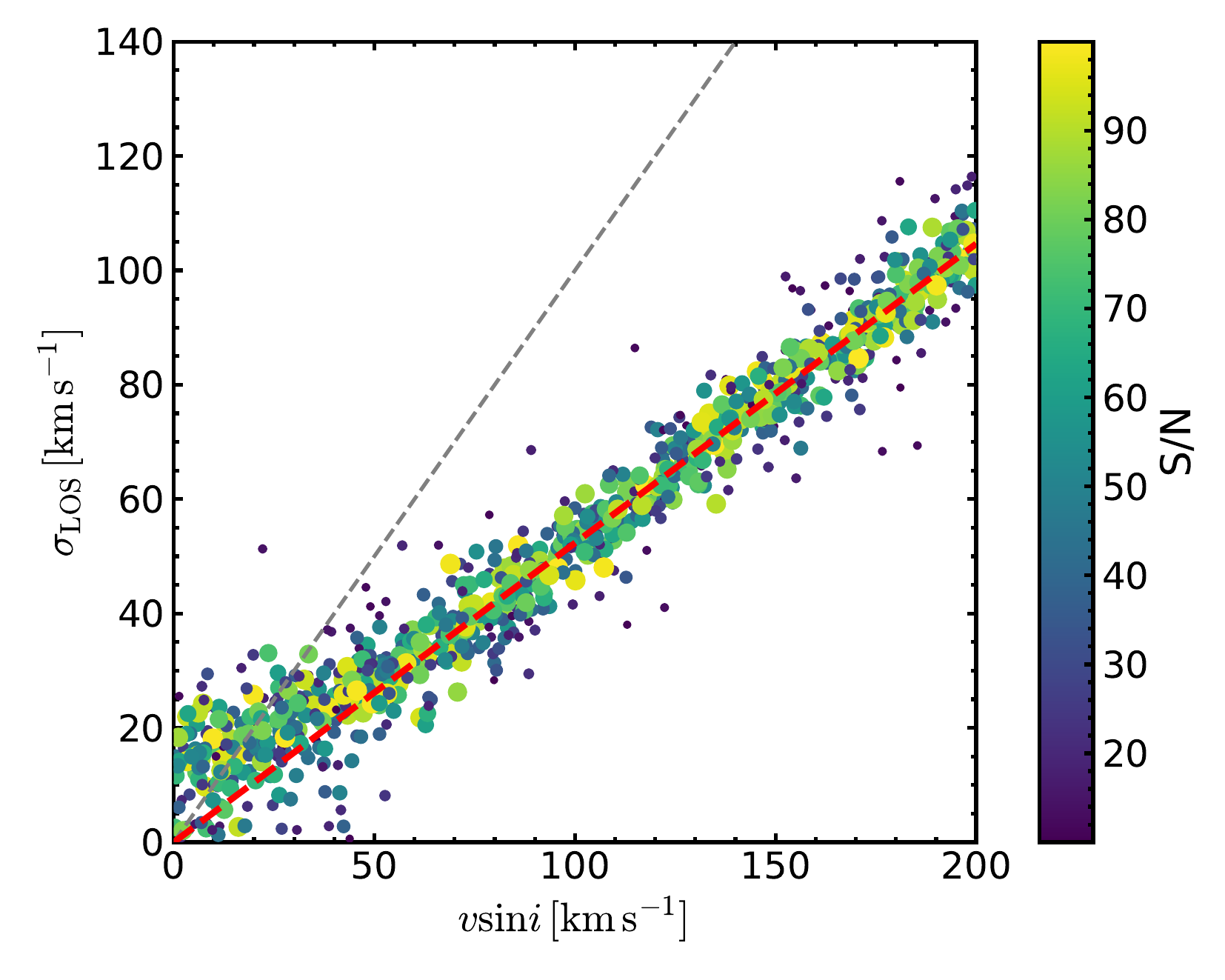}
    \caption{Relation between the input projected rotational velocity $v\sin i$ and the recovered line broadening $\sigma_{\rm LOS}$, obtained using mock MUSE spectra generated from an observed spectrum of the star HD~18769 (\textit{left panel}) and from a synthetic PHOENIX spectrum (\textit{right panel}). In both panels, different symbols sizes and colours indicate the S/N of the mock spectra. The thin black dashed lines show a one-to-one relation while the thick red dashed lines show the best-fit relation according to eq.~\ref{eq:sigma_vsini}. Note that the reason for the cut-off at low $v\sin i$ visible in the left panel is the intrinsic rotation of HD~18769 ($v\sin i\approx40\,{\rm km\,s^{-1}}$).}
    \label{fig:vsini_test}
\end{figure*}

We show the result of our simulation in the left panel of Fig.~\ref{fig:vsini_test}, where the line broadening $\sigma_{\rm LOS}$ determined by \textsc{Spexxy} is plotted as a function of the $v\sin i$ used to generate the spectra. We obtain a relation that is very close to being linear. The best fitting linear relation  that passes through the origin gives
\begin{equation}
 \sigma_{\rm LOS} = 0.55\times v\sin i\,.
 \label{eq:sigma_vsini}
\end{equation}
It is reassuring that no strong dependence on the S/N of the simulated spectra is observed. Only at the lowest values, ${\rm S/N} < 15-20$, the recovered $\sigma_{\rm LOS}$ are slightly too low, leading to $v\sin i$ values underestimated by $\sim5-10\,{\rm km\,s^{-1}}$ when using eq.~\ref{eq:sigma_vsini}. At higher S/N, the scatter around the relation is $\sim10\,{\rm km\,s^{-1}}$, confirming that we can measure $v\sin i$ to good accuracy in the simulated range from $40-200\,{\rm km\,s^{-1}}$. Furthermore, it is reassuring that no significant impact of the LSF or the initial stellar parameters on the results is observed.

We performed a second set of simulations in which we replaced the observed X-Shooter spectrum by a synthetic PHOENIX spectrum from the \textsc{G\"ottingen Spectral Library} \citep{2013A&A...553A...6H}, with stellar parameters resembling those of eMSTO stars in NGC~1846 ($T_{\rm eff}=7200\,{\rm K}$, $\log g=4.0$, $[{\rm M/H}]=-0.5$). As mentioned above, the reason for performing this second set of simulations was the intrinsic rotation of HD~18769, which for example prevented us from studying the recovery of $v\sin i$ in the regime $\lesssim 40\,{\rm km\,s^{-1}}$.

We set up and analysed the additional simulations in the same way as described above. The results of this process are shown in the right panel of Fig.~\ref{fig:vsini_test}. In comparison to the previous results, shown in the left panel of Fig.~\ref{fig:vsini_test}, we notice several characteristics. First, we obtain a very similar slope for the $\sigma_{\rm LOS}$-$v\sin i$ relation, $0.52$ instead of $0.55$ (cf. eq.~\ref{eq:sigma_vsini}). This illustrates that the ansatz of convolving the two rotation kernels as if they were Gaussians we made in setting up the first set of simulations does not lead to a deviant $v\sin i$ scale. Second, we find that for low simulated $v\sin i$, the relation flattens and we reach a plateau at $\sigma_{\rm LOS}\sim15\,{\rm km\,s^{-1}}$. We interpret this as evidence that our analysis is insensitive towards determining values of $v\sin i \lesssim40\,{\rm km\,s^{-1}}$.

The aforementioned flattening also implies that our measurement uncertainties become asymmetric at low $v\sin i$. In order to quantify this effect, we determined the 16th and 84th percentile of the distribution of recovered $v\sin i$ values for different values of input (i.e. true) $v\sin i$ (using a step size of $5\,{\rm km\,s^{-1}}$ in input $v\sin i$). We found that the uncertainties of our measured $v\sin i$ values become increasingly asymmetric below $100\,{\rm km\,s^{-1}}$ and are comparable to or larger than the actual measurements for $v\sin i\lesssim40\,{\rm km\,s^{-1}}$, again highlighting the sensitivity limit of our analysis. We incorporated this uncertainty calibration into our actual data.

\section{Serendipitous discovery of a planetary nebula in NGC~1846}
\label{app:nebula}

\begin{figure*}
    \centering
    \includegraphics[width=\textwidth]{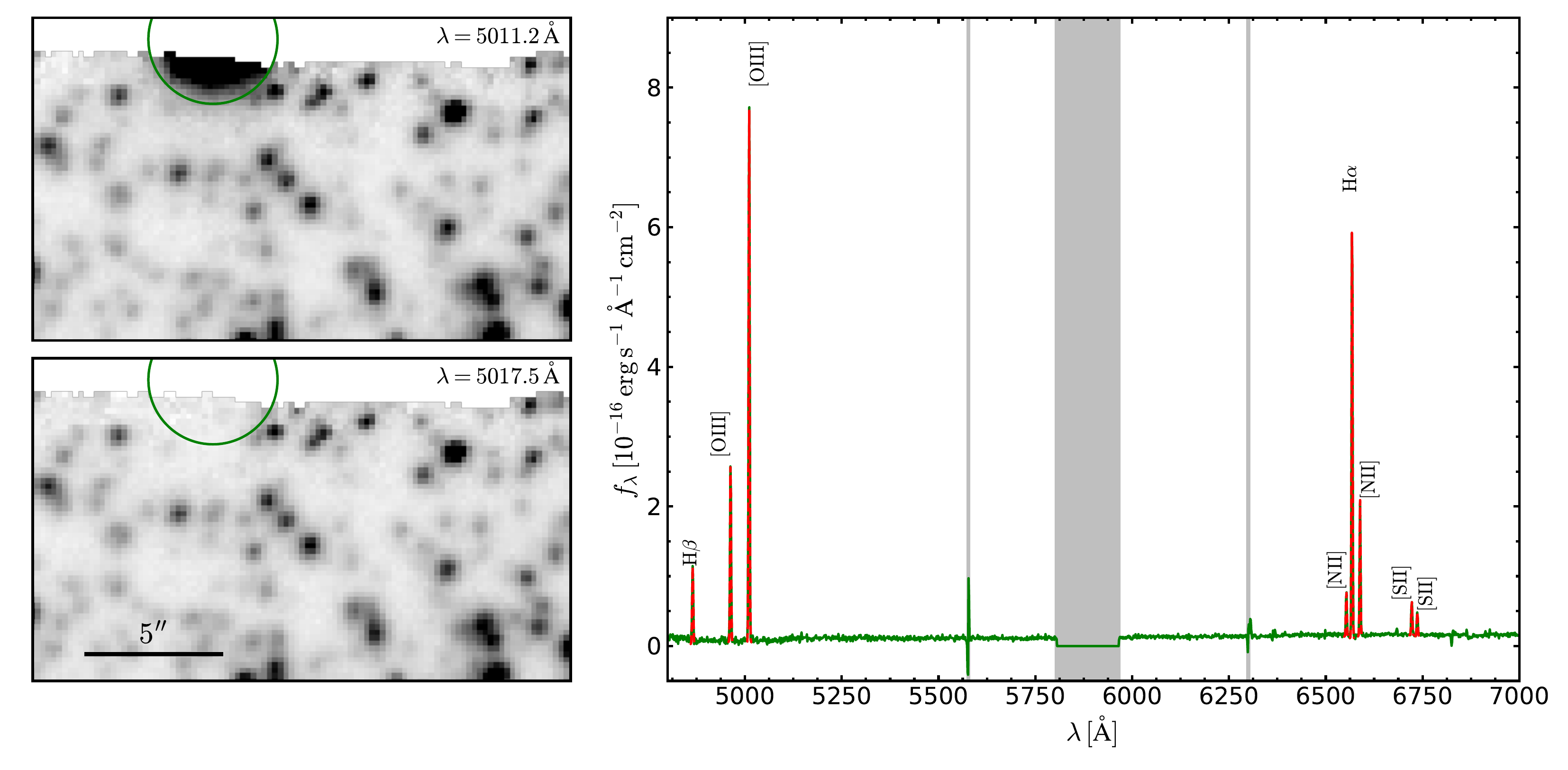}
    \caption{Discovery of a planetary nebula in NGC~1846. The left panels show an on-/off-band comparison for the $[\ion{O}{III}]5007$ line, with the nebula visible at the northern edge of the MUSE field of view in the on-band image. The right panel displays the spectrum extracted from the green circle visible in the left panels. Gaussian profiles have been fitted to the labelled emission lines. The grey-shaded areas mark the sodium laser gap and spectral regions affected by strong telluric emission.}
    \label{fig:nebula}
\end{figure*}

As illustrated by the discovery of a nova remnant in the Galactic globular cluster M22 (NGC~6656) by \citet{2019A&A...626A..69G}, MUSE is very efficient in uncovering faint nebular emission inside the $1\arcmin\times1\arcmin$ field of view. Here, we report the discovery of a planetary nebula in NGC~1846. As illustrated in Fig.~\ref{fig:nebula}, close inspection of our data revealed nebular emission at the northern edge of the MUSE pointing shown in Fig.~\ref{fig:colour_image}. Its spectrum, obtained via simple aperture extraction of the fluxes from all spaxels inside the green circle shown in Fig.~\ref{fig:nebula}, shows emission lines of hydrogen, oxygen, nitrogen, and sulphur (cf. right panel of Fig.~\ref{fig:nebula}). Based on the strong $\ion{O}{iii}$ lines, the weak $\ion{S}{ii}$ lines, and the ratio of the two visible $\ion{S}{ii}$ lines, we classify the object as a planetary nebula (PN).

Unfortunately, the MUSE field of view only covers a fraction of the PN, so its morphology and exact location remain unknown. We also inspected the HST images available for the location under question, but none of them clearly suggested the presence of the PN, only the F555W image appeared to show a marginal glow around this position. Based on the green circle included in Fig.~\ref{fig:nebula}, we estimate its central coordinates to $\alpha=05^{\rm h}\,07^{\rm m}\,36^{\rm s}78$ and $\delta=-67^{\circ}\,27\,\arcmin\,08\farcs8$. The catalogue of PNe in the Large Magellanic Cloud by \citet{2006MNRAS.373..521R} does not list any objects near that location. In particular, we stress that this PN is not to be confused with Mo-17, which was discovered $\sim80\arcsec$ south of NGC~1846 by \citet{1994A&AS..103..235M} and is likely associated with the cluster \citep{2013ApJ...762...65M}. Hence we consider the object as a PN that was previously unknown.

We further fitted Gaussian line profiles to the main emission lines visible in the right panel of Fig.~\ref{fig:nebula}. Their centroids yielded a radial velocity of $241.7\pm7.1\,{\rm km\,s^{-1}}$, which is in good agreement with the systematic velocity of NGC~1846. This makes it very likely that the PN is part of NGC~1846, instead of being a foreground or background object. When inserting the radial velocity and the distance to the cluster centre in our membership determination method (cf. Sec.~\ref{sec:membership}), we obtain a probability of $96\%$ that the PN is a cluster member.

\begin{figure}
    \centering
    \includegraphics[width=\columnwidth]{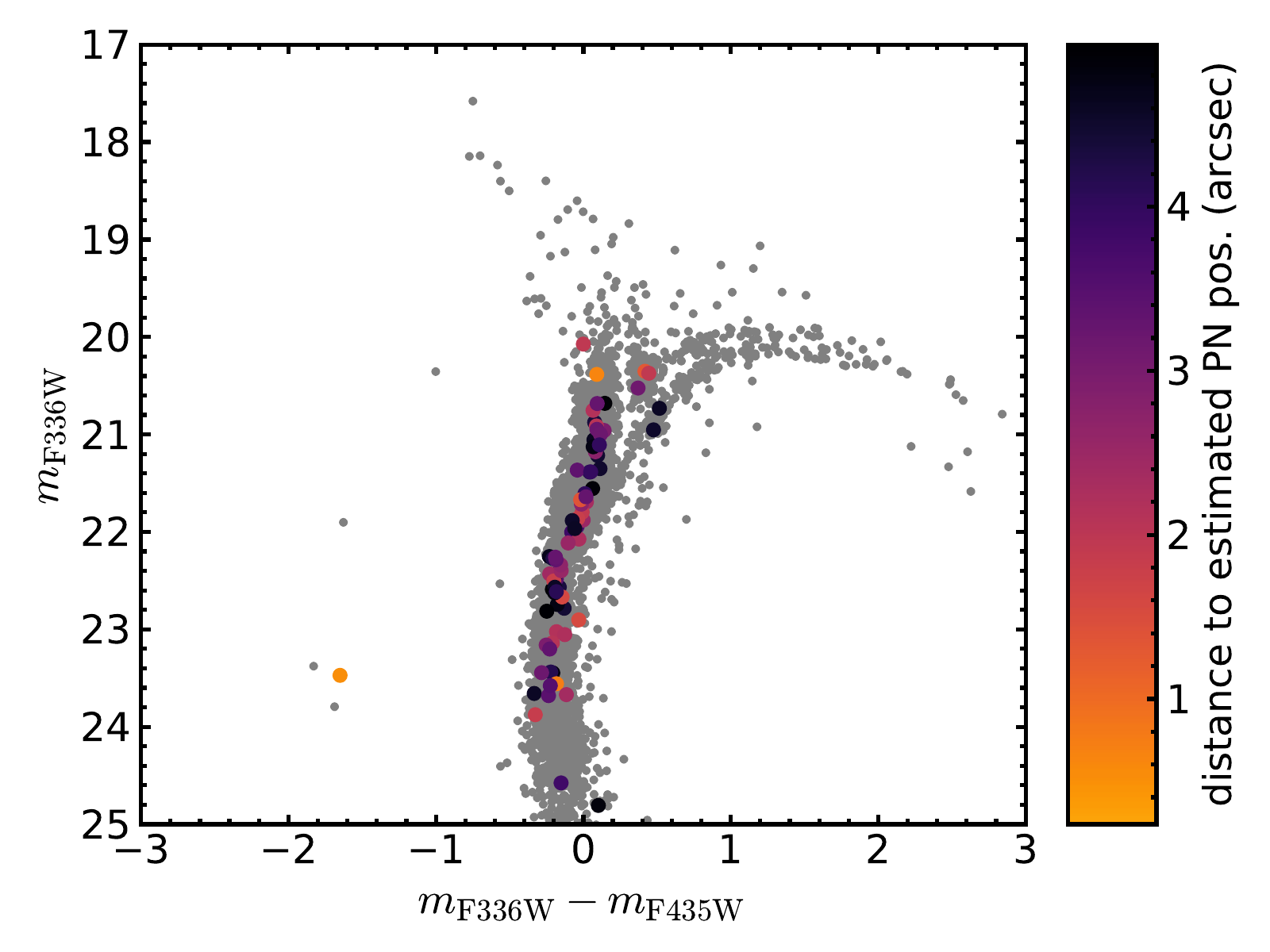}
    \caption{Near-UV colour-magnitude diagram of NGC~1846, created from the photometry presented in Sect.~\ref{sec:extraction}. Stars available in the HST catalogue within $5\arcsec$ of the estimated centre of the planetary nebula are highlighted according to their distance to said position.}
    \label{fig:nebula_nearby_stars}
\end{figure}

We also searched the HST photometry for a possible central star of the PN. However, within a distance of $5\arcsec$ to the estimated centre position, no obvious candidates that are bright enough and show very blue colours were detected (cf. Fig.~\ref{fig:nebula_nearby_stars}). We note the presence of one blue star (with $m_{\rm F336W}=23.4$ and $m_{\rm F435W}=25.1$), which is consistent with being a central star of a PN on the cooling track. However, given its faintness, the associated nebula should have already dissolved. It is also possible that the central star is strongly extincted.


\bsp	
\label{lastpage}
\end{document}